\documentclass[12pt]{article}
\usepackage[T1]{fontenc}
\usepackage[latin1]{inputenc}
\usepackage{a4wide}
\usepackage{amsmath}
\usepackage{color}
\usepackage{graphics}
\usepackage{amssymb}

\makeatletter

\providecommand{\LyX}{L\kern-.1667em\lower.25em\hbox{Y}\kern-.125emX\@}

\let\SF@@footnote\footnote
\def\footnote{\ifx\protect\@typeset@protect
    \expandafter\SF@@footnote
  \else
    \expandafter\SF@gobble@opt
  \fi
}
\expandafter\def\csname SF@gobble@opt \endcsname{\@ifnextchar[
  \SF@gobble@twobracket
  \@gobble
}
\edef\SF@gobble@opt{\noexpand\protect
  \expandafter\noexpand\csname SF@gobble@opt \endcsname}
\def\SF@gobble@twobracket[#1]#2{}
\makeatother

\begin{document}

\begin{center}
{\Large Finite Size Effects in Integrable Quantum Field Theories}\\
\vspace{.5cm}
{\large Francesco Ravanini}\\
\vspace{.3cm}
{\em I.N.F.N. - Sezione di Bologna, Via Irnerio 46 - 40126 Bologna,
Italy}\\ E-mail: ravanini@bologna.infn.it
\end{center}
\begin{abstract}
The study of Finite Size Effects in Quantum Field Theory allows the extraction
of precious perturbative and non-perturbative information. The use of scaling
functions can connect the particle content (scattering theory formulation) of
a QFT to its ultraviolet Conformal Field Theory content. If the model is 
integrable,
a method of investigation through a nonlinear integral equation equivalent to
Bethe Ansatz and deducible from a light-cone lattice regularization is 
available.
It allows to reconstruct the S-matrix and to understand the locality properties
in terms of Bethe root configurations, thanks to the link to ultraviolet CFT
guaranteed by the exact determination of scaling function. This method is 
illustrated
in practice for Sine-Gordon / massive Thirring models, clarifying their 
locality
structure and the issues of equivalence between the two models. By restriction
of the Sine-Gordon model it is also possible to control the scaling functions
of minimal models perturbed by \( \Phi _{(1,3)} \).\vspace{1cm}
\end{abstract}

\section{Introduction}

\begin{figure}[h]
{\par\centering 
\resizebox*{0.7\textwidth}{0.35\textheight}{\includegraphics{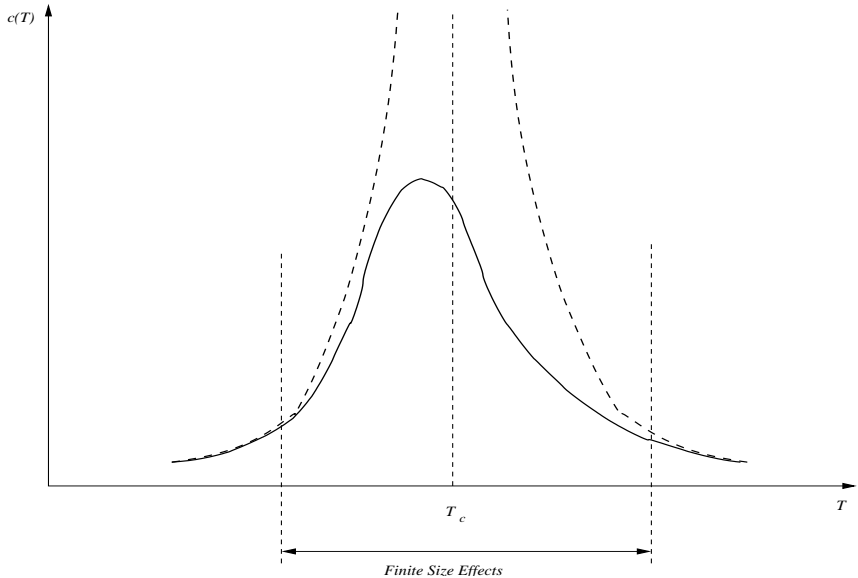}} 
\par}

\caption{\label{fig: cT}{\small Specific Heat near a phase transition point in 
infinite
volume (dashed line) and in finite volume (solid line)}\small }
\end{figure}

Finite Size Effects (FSE) are widely recognized to play an important role in
modern Statistical Mechanics and Quantum Field Theory (QFT). From a statistical
point of view, it is known that no phase transitions take place in a finite
volume system. For example, the specific heat \( c(T) \), that is divergent
at the critical point in infinite volume, looses it divergence if the system
has finite size; one observes only a rounded peak in the plot of \( c(T) \)
versus \( T \) (see fig.\ref{fig: cT}). Moreover, there is only an interval
around critical temperature \( T_{c} \) where the FSE are relevant. Out of
this interval, they are negligible (because only near \( T_{c} \) the 
correlation
length is comparable with the size of the system). The interesting fact is that
specific heat (and other critical quantities) have a scaling behaviour (i.e.
varying the size \( L \)) that is fixed by the (infinite size) critical 
exponents
(see \cite{Christe_Henkel}). This is a general fact: as argued in 
\cite{cardy86},
the UV behaviour of the scaling functions (see later) is fixed by the conformal
dimensions of the operators that belong to the universality class of the 
critical
point (i.e. the CFT describing the critical point of the statistical system).

\begin{figure}[h]
{\par\centering 
\resizebox*{0.7\textwidth}{0.35\textheight}{\includegraphics{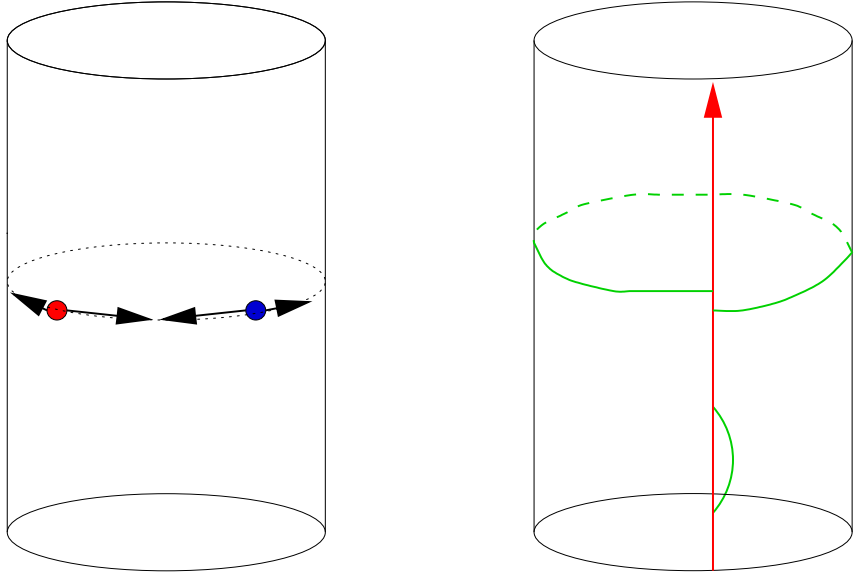}} 
\par}

\caption{{\small Radiative corrections to the self energy of a particle in 
finite volume}
\label{fig: cyl}}
\end{figure}

Also in QFT interesting phenomena appear. If the space-time geometry is a 
cylinder
of circumference \( L \), Casimir effects change the energy of a two body 
interaction,
because particles interact in the two possible directions, as shown in fig.
\ref{fig: cyl}. New radiative corrections to the self-energy of a propagating
particle may appear because of the closed geometry. 
For example, a massive particle propagating in a finite 1+1 dimensional 
space-time,
with periodic boundary conditions in the space direction and an infinitely 
extended
time direction will have its propagator renormalized by the contributions of
radiative corrections, as depicted in the second cylinder of fig.\ref{fig: 
cyl}.
While on an infinite space only local virtual emissions are allowed, on a 
cylindrical
geometry one can conceive virtual emissions that travel around the whole 
cylinder
before coming back to the bare particle. Of course they are exponentially 
depressed
if compared to the traditional local emissions, therefore, if the cylinder is
large, their contributions to the self-energy of the propagator is negligible.
However, if the cylinder circumference \( L \) is small enough, there is room
left by Heisenberg principle for such virtual object to travel around the world
and come back from the other side, thus giving new contributions that modify
the self-energy.

Lüscher \cite{luscher_1,luscher_2} has estimated the relevance of such a 
phenomenon
by computing the corrections to the mass of the particle due to FSE. For a 
massive
self-interacting boson with \( \phi ^{4} \) potential in 3+1 dimensions, they
are represented by\[
\Delta M(L)=-\frac{1}{16\pi ^{2}}\frac{3}{mL}\int dy\, 
e^{-\sqrt{m^{2}+y^{2}}}F(iy)+O(e^{-\sqrt{\frac{3}{2}}mL})\]
 \( m \) is the physical mass of the particle in infinite volume; \( F \)
is the analytic continuation of the forward elastic scattering amplitude \( S 
\)
with incoming momenta \( p_{1},p_{2} \) and outcoming momenta \( p_{3},p_{4} \)
\[
F(\nu )=S(p_{1},p_{2}|p_{3},p_{4})\]
where \( \nu =\frac{\omega (p_{1})\omega (p_{2})-p_{1}\cdot p_{2}}{m} \) and
\( \omega (p)=\sqrt{m^{2}+\vec{p}^{2}} \). Saddle point integration peaks a
contribution from around \( y=0 \), so that the result is \[
\Delta M\propto \frac{e^{-mL}}{L\sqrt{mL}}+...\]

A very useful application of this estimate shows up in lattice QFT 
calculations.
For example consider a 4 dimensional lattice, of lattice spacing \( a \), on
which you want to perform Montecarlo simulations of QCD. The lattice sizes are
\( T\times L\times L\times L \), with, say, \( L=20a \). In lattice QFT, the
discretization of space is to be intended as a regulator, \emph{i.e.} it is
equivalent to the introduction of a momentum cutoff \( \Lambda _{cut}\approx 
\frac{1}{a} \)
which must be much larger than the typical hadron masses, if we want to get
reliable data from the lattice. For example, we can choose \( a<0.1 \) fm (1
fm = \( 10^{-13} \) cm); then \( L=20a=\, 2 \) fm, which is comparable with
the proton radius (Compton wavelength). The proton in such a lattice is 
{}``squeezed{}'',
which means that the FSE are relevant. One way to overcome this difficulty is
to build larger and larger computers, which, as history of the last 20 years
indicates, goes to introduce other sorts of difficulties of financial, 
technical
and also principal nature. The other possibility, at the cost of a pencil, is
to be able to compute the FSE scaling of quantities related to the proton, from
which we can extract physical data even from such small lattices. Of course
nobody is nowadays able to estimate with accuracy the \( L \) dependence of
quantities in a formidable problem like QCD, but progresses have been made in
two dimensional QFT's, some of them sharing interesting phenomena with QCD or
other interesting realistic 4 dimensional QFT's. In particular, there are 
methods
to exactly calculate the \( L \) dependence of physical quantities in the class
of \emph{Integrable} two dimensional QFT. This will be the main topic of the
present review.

Let us continue to analyze this lattice QFT example. Finite Size with periodic
boundary conditions implies quantization of momenta in units of order \( 
\Delta p=\frac{2\pi }{L}\geq 600 \)
MeV. Pions in such a lattice (\( m_{\pi }=139 \) MeV) have virtual processes
strongly affected by FSE. This illustrates a bit more numerically the 
observation
about virtual processes made above.

This discussion leads us to some important observations:

\begin{itemize}
\item As momenta are quantized, the energy spectrum for finite \( L \) is 
discrete
\item FSE do not depend strongly on the lattice spacing \( a \), they depend 
on the
scale \( L\gg a \) and become irrelevant when \( L\, \gg  \) Compton wavelength
of the lightest particle (of mass \( m \)). Therefore the physical quantities
may be expressed, up to overall factors encoding their dimensionality, as 
dimensionless
functions of the dimensionless scale \( l=mL \). To such objects the usual
denomination of \emph{scaling functions} is given.
\item As \( L\gg a \), FSE are relevant to renormalized QFT. \( l \) can play 
the
role of dimensionless scale parameter of a sort of Renormalization Group flow
as we go to illustrate by the following heuristic arguments:

\begin{itemize}
\item the limit \( l\rightarrow 0 \) means or \( L\rightarrow 0 \) i.e. the 
cylinder
shrinks to a line, and there is no spatial dimension anymore, which is an 
uninteresting
unphysical situation, or \( m\rightarrow 0 \), which means that the lightest
particle mass becomes irrelevant compared to the energies in play in the 
system,
which is exactly the setup of the ultraviolet (UV) regime. Therefore, the limit
\( l\rightarrow 0 \) reproduces the UV regime of the model.
\item the limit \( l\rightarrow \infty  \) means or \( m\rightarrow \infty  \),
i.e. the situation where the masses of the particles are much greater than the
scattering transferred momenta, which defines the infrared (IR) region, or \( 
L\rightarrow \infty  \),
i.e. the reconstruction of the QFT on the infinite volume. Note that these two
limits are not in contradiction: actually the theory on the whole infinite 
volume
is described in terms of the S-matrix, which is an intrinsically IR object,
in the sense that it describes the scattering of asymptotic particles without
looking at the details of the interaction at short range.
\end{itemize}
\item Both the UV and IR limits thus identified are fixed points of the QFT 
where
the Callan - Symanzyk beta function annullates. They are therefore scale 
invariant
points. Scale invariance in relativistic QFT often implies conformal 
invariance.
Therefore the UV and IR points of the flow obtained by varying \( l \) from
0 to \( \infty  \) are to be identified with Conformal Field Theories (CFT). 

\begin{itemize}
\item In the case of massive QFT, the IR CFT is trivial. The CFT description 
is useful
in this case only at the UV point, where the theory can be seen as a relevant
perturbation of a given CFT. At IR instead, the most effective description is
in terms of the S-matrix.
\item In the case of massless (but not scale invariant) theories, both the UV 
and
IR points are nontrivial CFT's, and the scaling functions interpolate between
the physical quantities of these two CFT's.
\end{itemize}
\end{itemize}

\section{Excursus on the Renormalization Group in 1+1 dimensions}

In this section we quickly recall the main properties of a two-dimensional QFT
under scale transformations. In general a QFT is not invariant under scale 
transformations,
due to the presence of dimensionful parameters, appearing already at the 
classical
level or generated by the renormalization procedure. To investigate this 
phenomenon
it is useful to think in terms of the so called \emph{(Wilson) Space of 
Actions}
\cite{Wilson,Ma}. Let us suppose that the action \( S=\int d^{D}x\, {\cal L} \)
of a theory depends on a set of fundamental fields and their derivatives, and
on a set of dimensionless coupling constants \( \vec{g}\equiv 
(g_{1},...,g_{n}) \).
The variation of the Lagrangian density \( {\cal L}(g_{1},...,g_{n}) \) under
scale transformations \( x\rightarrow x+dt\, x \) can be seen as a 
transformation
of the coupling constants \( \vec{g}\rightarrow \vec{g}' \). Scale invariance
in this context means that under a scale transformation the point \( \vec{g} \)
remains unchanged. Scale transformations induce a sort of \emph{motion} in the
\( n \)-dimensional space of couplings. Such a space is called usually the
Space of Actions. A trajectory in this space is a function \( \vec{g}(t) \)
(running coupling constants). The variation of the Lagrangian is \( {\cal 
L}(t+dt)={\cal L}(t)+\partial _{\mu }J^{\mu }dt \)
where \( J^{\mu }=x_{\nu }T^{\mu \nu } \) is the Nöther current associated
to the scale transformation. The object \( \partial _{\mu }J^{\mu }=T_{\mu 
}^{\mu }\equiv \Theta  \)
is the trace of the stress energy tensor (we shall always consider in the 
following
a symmetrized improved stress energy tensor \cite{Callan-Coleman-Jackiw}) and
scale invariance, where \( J^{\mu } \) satisfies a continuity equation \( 
\partial _{\mu }J^{\mu }=0 \),
implies \( \Theta =0 \). For a general trajectory \( \vec{g}(t) \) the stress
energy tensor is not traceless. The \emph{Callan - Symanzyk beta function} is
defined as\begin{equation}
\label{beta}
\beta _{i}(\vec{g})\equiv \frac{dg_{i}}{dt}
\end{equation}
 and can be interpreted as a sort of \emph{velocity field} of the theory in
the action space. Let us consider one coupling for simplicity. Imagine to be
at a certain point in the space of actions where the action of the theory is
characterized by a parameter \( g_{0} \). Consider this as a sort of initial
condition, and study the behaviour of the theory as \( t \) increases.

\begin{itemize}
\item if \( \beta (g_{0})>0 \) then \( g(t) \) is increasing in a region near 
\( g_{0} \);
\item if \( \beta (g_{0})<0 \) then \( g(t) \) is decreasing in a region near 
\( g_{0} \).
\end{itemize}
Points \( g^{*} \) where \( \beta (g^{*})=0 \) are called \emph{fixed points}.
They divide the space of actions (one-dimensional in this example) in distinct
regions of the \( g \) parameter. The running coupling constant keeps confined
in one of these regions. Independently of the exact value of \( g_{0} \), in
a given region, the asymptotic value of \( g(t) \) for \( t\rightarrow \pm 
\infty  \)
is always given by the closest values of \( g \) where \( \beta (g)=0 \).
There are two types of fixed points:

\begin{itemize}
\item a fixed point \( g^{*} \) is \emph{infrared} (IR) \emph{}if, starting 
from
a value \( g_{0} \) of the coupling near it, one reaches it for \( 
t\rightarrow +\infty  \).
(\( \beta '(g^{*})<0 \)).
\item a fixed point \( g^{*} \) is \emph{ultraviolet} (UV) if, starting from a 
value
\( g_{0} \) of the coupling near it, one reaches it for \( t\rightarrow 
-\infty  \).
(\( \beta '(g^{*})>0 \)).
\end{itemize}
In the case with more parameters all the beta functions have to annullate 
simultaneously
in a fixed point. Given the initial conditions \( \vec{g}=\vec{g}_{0} \) there
exists a unique trajectory, as the equation (\ref{beta}) is a first order 
linear
differential equation. The trajectory \( \vec{g}(t) \) is often referred in
the literature as the \emph{Renormalization Group (RG) flow.}

The knowledge of the beta function is enough to reconstruct the behaviour of
the theory in a surrounding of the point \( g_{0} \) i.e. to have a complete
description of the tangent space to the space of actions in the neighborhood
of \( g_{0} \). In such tangent space, the variation of the Lagrangian can
be expressed as a linear combination of a base of fields identified as the 
derivatives
of the Lagrangian with respect to the coupling constants \( g_{i} \) \[
\Phi _{i}=\frac{\partial {\cal L}}{\partial g_{i}}\]
 The trace of the stress energy tensor is a field living in such tangent 
space\[
\Theta (x)=\frac{d{\cal L}}{dt}=\sum _{i}\frac{\partial {\cal L}}{\partial 
g_{i}}\frac{dg_{i}}{dt}=\sum _{i}\beta _{i}(\vec{g})\Phi _{i}(\vec{g})\]
as it is a linear combination of the fields \( \Phi _{i} \) with coefficients
given by the components of the beta function vector. It is then obvious that
\( \Theta =0 \) when and only when \( \beta _{i}(\vec{g})=0 \), which means
that the fixed points of a QFT are scale invariant.

We go now to get an equation describing the variation of the \( N \) point
correlation functions of generic fields \( A_{i}(x) \) of the theory \( 
\langle X\rangle \equiv \langle A_{1}(x_{1}),...,A_{N}(x_{N})\rangle  \)
along the RG flow. It is known in the literature as the Callan-Symanzyk 
equation.
We present here a derivation inspired by A. Zamolodchikov \cite{Zam86}. The
variation of correlation functions under scale transformation can be described
in two equivalent ways

\begin{enumerate}
\item apply the variation to all fields in the correlator\[
\begin{array}{ccc}
\delta \langle X\rangle  & \equiv  & \delta \langle 
A_{1}(x_{1}),...,A_{N}(x_{N})\rangle =\delta \int {\cal D}\varphi \, 
A_{1}...A_{N}\, e^{-S[\phi ]}\\
 & = & \delta \int {\cal D}\varphi \, \left[ \sum _{k}A_{1}...\delta 
A_{k}...A_{N}\, e^{-S}-A_{1}...A_{N}\, e^{-S}\delta S\right] 
\end{array}\]
 The variation of a generic field \( A_{k} \) under scale transformation is
given by the global dilation of coordinates and by the intrinsic variation of
the field due to its ingeneering (classical) dimension \( D_{k} \)\[
\delta A_{k}=dt(x^{\mu }_{k}\partial _{\mu }^{k}+D_{k})A_{k}\]
 The variation of the action is actually given by the insertion of the trace
of the stress energy tensor into the correlator\[
\delta S=\int d^{2}y\Theta (y)\]
 Therefore\[
\delta \langle X\rangle =dt\left( \sum _{k}(x_{k}^{\mu }\partial _{k}^{\mu 
}+D_{k})\langle X\rangle -\int d^{2}y\langle X\Theta (y)\rangle \right) \]
 
\item alternatively one can vary the fields and the action with respect to the 
couplings
\( g_{i} \)\[
\delta \langle X\rangle =\sum _{i}\beta _{i}(\vec{g})\frac{\partial }{\partial 
g_{i}}\langle X\rangle \]
 but \[
\frac{\partial }{\partial g_{i}}\langle X\rangle =\sum _{k}\langle 
A_{1},...,\frac{\partial A_{k}}{\partial g_{i}},...,A_{N}\rangle -\int \langle 
X\Phi _{i}(x)\rangle d^{2}x\]
 Be careful, in these formulae, not to confuse summations over \( i \), that
run in the tangent to the space of actions and summations over \( k \) that
run over the points \( x_{k} \) in the correlator. Putting together the pieces,
one gets\[
\delta \langle X\rangle =\sum _{i}\sum _{k}\langle A_{1},...,\frac{\partial 
A_{k}}{\partial g_{i}},....A_{N}\rangle -\int d^{2}y\langle X\Theta (y)\rangle 
\]

\end{enumerate}
Combining the two methods one arrives at the so called Callan - Symanzyk 
equation
for the 1+1 dimensional RG of QFT\[
\left[ \sum _{k}\left( x_{k}^{\mu }\frac{\partial }{\partial x_{k}^{\mu 
}}+\Gamma _{k}(\vec{g})\right) -\sum _{i}\beta _{i}(\vec{g})\frac{\partial 
}{\partial g_{i}}\right] \langle X\rangle =0\]
where \( \Gamma _{k}A_{k}(x_{k})=\left( D_{k}+\sum _{i}\beta 
_{i}(\vec{g})\frac{\partial }{\partial g_{i}}\right) A_{k}(x_{k}) \)
is the so called \emph{anomalous dimension} of the field \( A_{k} \).

In particular the dimensions of the perturbing fields \( \Phi _{i} \) are given
by\[
\begin{array}{ccc}
\Gamma (\vec{g})\Phi _{i}(x) & = & \sum _{j}\gamma _{i}^{j}(\vec{g})\Phi 
_{j}(x)=\left( D_{k}+\frac{d}{dt}\right) \frac{\partial L}{\partial g_{i}}\\
 & = & \frac{\partial }{\partial g_{i}}(2L+\Theta )=\sum _{j}\left( 2\delta 
_{i}^{j}-\frac{\partial \beta _{j}}{\partial g_{i}}\right) \Phi _{j}(x)
\end{array}\]
 Applying this to the trace of the stress energy tensor \( \Theta =\sum 
_{i}\beta _{i}\Phi _{i} \)
one gets \( \Gamma \Theta =2\Theta  \). This, combined with Lorentz covariance,
implies that also the other components of the stress energy tensor have the
same anomalous dimension 2. This is also the ingeneering dimension of the 
stress
energy tensor, which means that the stress energy tensor does not develop an
anomalous dimension and keeps its classical dimension also at the quantum 
level.
This is a fact common to all conserved currents.

The most exciting result known in the study of two-dimensional Renormalization
Group is undoubtedly the Zamolodchikov \emph{c-theorem} \cite{Zam_cth}. Here
we do not reproduce the entire argument leading to this theorem (see the 
original
paper \cite{Zam_cth} for a demonstration). We only offer the statement of the
theorem in its full generality.

\textbf{c-theorem}: In any \emph{unitary} QFT\( _{2} \) a function \( 
c(\vec{g}) \)
exists, such that:

\begin{enumerate}
\item it is decreasing along the RG flow: \( \frac{dc}{dt}\leq 0 \);
\item it is stationary in the fixed points \( \vec{g}=\vec{g}^{*} \), i.e. \( 
\left. \frac{dc}{dt}\right| _{\vec{g}^{*}}=0 \)
if and only if \( \vec{g}^{*} \) is a fixed point of the RG flow, i.e. \( 
\beta _{i}(\vec{g}^{*})=0 \).
\item in such fixed points \( \vec{g}^{*} \), the function \( c(\vec{g}) \) 
fixes
completely the two-point correlators of the stress energy tensor. Introducing
the coordinates \[
z=x_{1}+ix_{2}\qquad \bar{z}=x_{1}-ix_{2}\]
 and the notation \( T=T_{zz} \), \( \bar{T}=T_{\bar{z}\bar{z}} \), \( \Theta 
=T_{z\bar{z}}=T_{\bar{z}z}=T_{\mu \mu } \)
one has \( \Theta =0 \) because the fixed point is scale invariant 
and\begin{equation}
\label{TTcorr}
\langle T(z,\bar{z})T(0,0)\rangle =\frac{c/2}{z^{4}}\qquad \langle 
\bar{T}(z,\bar{z})\bar{T}(0,0)\rangle =\frac{c/2}{\bar{z}^{4}}\qquad \langle 
T\bar{T}\rangle =0
\end{equation}
 where \( c=\lim _{\vec{g}\rightarrow \vec{g}^{*}}c(\vec{g}) \). The constant
\( c \) is often called \emph{central charge} and is a characteristic of the
theory at the fixed point. 
\end{enumerate}
As we have seen that theories are RG flows from UV to IR fixed points when \( 
t \)
goes from \( -\infty  \) to \( +\infty  \). This theorem means that, for 
unitary
theories \[
c_{UV}\geq c_{IR}\]
which is the {}``simplified{}'' form frequently mentioned in the literature.

\section{Conformal Field Theory}

In the previous section we have seen that the RG fixed points correspond to
values of the parameters of a theory where scale invariance is guaranteed. 
Scale
invariance implies under quite general assumptions a larger invariance, known
as conformal invariance. In this section we summarize the main results in the
treatment of conformal invariant theories (CFT) in two-dimensions, as developed
in the fundamental work \cite{bpz}.

\subsection{The Virasoro algebra and the Hilbert space}

In two dimensions any analytic transformation of co-ordinates \( 
z=x_{1}+ix_{2} \)
and \( \bar{z}=x_{1}-ix_{2} \) is a conformal transformation. The improved
stress energy tensor obeys the continuity equation \( \partial _{\mu }T^{\mu 
\nu }=0 \)
and is symmetric \( T^{\mu \nu }=T^{\nu \mu } \). If we add the requirement
of scale invariance, i.e. of null trace \( T^{\mu \mu }=\Theta =0 \), we end
up, in two dimensions, to have only two independent components \( 
T=T_{zz}=\frac{1}{4}(T^{11}-T^{22}+2iT^{12}) \)
and \( \bar{T}=T_{\bar{z}\bar{z}}=\frac{1}{4}(T^{11}-T^{22}+2iT^{12}) \), for
which the continuity equations read\[
\partial _{\bar{z}}T(z,\bar{z})=\partial _{z}\bar{T}(z,\bar{z})=0\]
 expressing that \( T \) and \( \bar{T} \) are functions only of \( z \)
and \( \bar{z} \) respectively: \( T=T(z) \) and \( \bar{T}=\bar{T}(\bar{z}) 
\).
The theory separates into two {}``non-interacting{}'' parts with dependence
on \( z \) and \( \bar{z} \) respectively. In what follows we shall concentrate
on the \( z \) part, although all statements can be repeated for the \( 
\bar{z} \)
part too.

The stress-energy tensor is the generator of conformal transformations, in the
sense that if we take its Laurent expansion \begin{equation}
\label{Tmodes}
T(z)=\sum _{n\in \mathbb {Z}}\frac{L_{n}}{z^{n+2}}\qquad {\rm i.e.}\qquad 
L_{n}=\oint _{0}\frac{dz}{2\pi i}T(z)z^{n+1}
\end{equation}
 then the mode \( L_{n} \) generates the infinitesimal conformal transformation
\( z\rightarrow z+\varepsilon z^{n+1} \) (analogous expansion holds for \( 
\bar{T}(\bar{z}) \)
with modes \( \bar{L}_{n} \)). In particular \( L_{-1} \) generates the 
translations
in the \( z \) direction, \( L_{0}+\bar{L}_{0} \) generates the dilations,
\( L_{0}-\bar{L}_{0} \) the rotations. One can write down conformal Ward 
identities,
that, once integrated, give the operator product expansion (OPE) for any field
with the stress-energy tensor. In particular the OPE of \( T \) with itself
reads as \begin{equation}
\label{TT}
T(z)T(w)=\frac{c/2}{(z-w)^{4}}+\frac{2T(w)}{(z-w)^{2}}+\frac{\partial 
T(w)}{z-w}+\mbox {regular\, terms}
\end{equation}
 The number \( c \) is called \emph{conformal anomaly} or often \emph{central
charge} and plays a central role in the following. Eq.(\ref{TT}) implies for
the modes (\ref{Tmodes}) the Virasoro (\( {\cal V} \)) algebra \begin{equation}
\label{virasoro}
[L_{m},L_{n}]=(m-n)L_{m+n}+\frac{c}{12}(m^{3}-m)\delta _{m+n,0}
\end{equation}
 where the constant \( c \) appears in the central extension term. The algebra
of the full conformal group (do not forget the \( \bar{z} \) part) is then
\( {\mathcal{V}}\otimes {\bar{\mathcal{V}}} \). We shall use the so called
\emph{radial quantization}, where the time is taken as \( \log |z| \). In this
description \( L_{0}+\bar{L}_{0} \) is the Hamiltonian, and its eigenvalues
must be bounded by below. All the states of a CFT must lie in some irreducible
representation of the algebra \( {\mathcal{V}}\otimes {\bar{\mathcal{V}}} \).
The representations we need must have \( L_{0} \) and \( \bar{L}_{0} \) 
eigenvalues
bounded by below, i.e. they must contain a so called \emph{highest weight 
state}
(HWS) \( |\Delta \rangle  \) for which \begin{equation}
\label{hws}
L_{0}|\Delta \rangle =\Delta |\Delta \rangle \qquad ,\qquad L_{n}|\Delta 
\rangle =0\qquad ,\qquad n>0
\end{equation}
 These representations are known as \emph{highest weight representations} 
(HWR).
The irreducible representations of \( {\cal V} \) are labelled by two numbers;
for HWRs these are \( c \) and \( \Delta  \). We shall denote the HWRs of
\( {\mathcal{V}} \) by \( {\mathcal{V}}_{c}(\Delta ) \). For a given theory,
\( c \) is fixed by eq.(\ref{TT}), then the Hilbert space \( {\mathcal{H}} \)
of the theory is built up of all possible representations \( 
{\mathcal{V}}_{c}(\Delta ) \)
at fixed \( c \), each one with a certain multiplicity: \begin{equation}
\label{hilbert}
{\mathcal{H}}=\bigoplus _{\Delta ,\bar{\Delta }}{\mathcal{N}}_{\Delta 
,\bar{\Delta }}{\mathcal{V}}_{c}(\Delta )\otimes 
{\bar{\mathcal{V}}}_{c}(\bar{\Delta })
\end{equation}
 If a certain \( {\mathcal{V}}_{c}(\Delta )\otimes 
{\bar{\mathcal{V}}}_{c}(\bar{\Delta }) \)
does not appear, then simply \( {\mathcal{N}}_{\Delta ,\bar{\Delta }}=0 \).
The numbers \( {\mathcal{N}}_{\Delta ,\bar{\Delta }} \) count the multiplicity
of each representation in \( {\mathcal{H}} \), this implies they must always
be non negative integers. They are not fixed by conformal invariance. 
Constrains
on them arise from other physical requirements such as locality \cite{dotfat}
or modular invariance \cite{cardy86}. 

From eq.(\ref{TT}) one easily computes the two point correlator of the stress
energy tensor, and comparing with eq.(\ref{TTcorr}) it turns out that \( c \)
is exactly the value of the \( c(\vec{g}) \) function of the c-theorem at this
particular fixed point.

Any state \( |s\rangle \equiv |\Delta ,\bar{\Delta }\rangle  \) in the theory
can be put in 1 to 1 correspondence with a field through the formula \( 
|s\rangle =A_{s}(0,0)|0\rangle  \),
where the vacuum \( |0\rangle  \) is projective (i.e. \( 
L_{0},\bar{L}_{0},L_{\pm 1},\bar{L}_{\pm 1} \))
invariant. In particular the HWS (\ref{hws}) correspond to some fields \( \phi 
_{\Delta ,\bar{\Delta }}(z,\bar{z}) \)
that transform under the conformal group as \begin{equation}
\label{primary}
\phi _{\Delta ,\bar{\Delta }}(z,\bar{z})=\left( \frac{\partial z'}{\partial 
z}\right) ^{\Delta }\left( \frac{\partial \bar{z}'}{\partial \bar{z}}\right) 
^{\bar{\Delta }}\phi _{\Delta }(z',\bar{z}')
\end{equation}
 They are called \emph{primary fields}. Their OPE with the stress-energy tensor
is given by \begin{equation}
\label{Tphi}
T(z)\phi _{\Delta ,\bar{\Delta }}(w,\bar{w})=\frac{\Delta \phi _{\Delta 
,\bar{\Delta }}(w,\bar{w})}{(z-w)^{2}}+\frac{\partial _{w}\phi _{\Delta 
,\bar{\Delta }}(w,\bar{w})}{z-w}+\mbox {regular\, terms}
\end{equation}
 Applying this formula to the vacuum and using (\ref{Tmodes}) we go back to
(\ref{hws}). A basis for all the other states of the theory (called 
\emph{secondaries})
can be obtained by applying strings of \( L_{n},n<0 \) to \( |\Delta \rangle  
\).
The commutation relations imply \begin{equation}
\label{secondaries}
L_{0}L_{n}^{k}|\Delta \rangle =(\Delta +nk)L_{n}^{k}|\Delta \rangle 
\end{equation}
 Therefore \( L_{0} \) eigenvalues organize the space \( 
{\mathcal{V}}_{c}(\Delta ) \)
(often called a \emph{module}) so that the states lie on a {}``stair{}'' whose
\( N \)-th step (called the \( N \)-th \emph{level}) has \( L_{0}=\Delta +N 
\)\begin{equation}
\label{levels}
\begin{array}{ccc}
\mbox {states} & \mbox {level} & L_{0}\\
...... & ... & ...\\
L_{-3}|\Delta \rangle \, ,\, L_{-2}L_{-1}|\Delta \rangle \, ,\, 
L_{-1}^{3}|\Delta \rangle  & 3 & \Delta +3\\
L_{-2}|\Delta \rangle \, ,\, L_{-1}^{2}|\Delta \rangle  & 2 & \Delta +2\\
L_{-1}|\Delta \rangle  & 1 & \Delta +1\\
|\Delta \rangle  & 0 & \Delta 
\end{array}
\end{equation}
 All the fields corresponding to the HWR \( {\mathcal{V}}_{c}(\Delta ) \) are
said to be in the \emph{conformal family} \( [\phi _{\Delta }] \) generated
by the primary field \( \phi _{\Delta } \).

\subsection{Unitarity}

We said that for a given theory, \( c \) is fixed by eq.(\ref{TT}). Conversely,
we can ask how many theories are there at a certain value of \( c \). The 
answer
to this question for all \( c \)'s is what is called \emph{classification}
of CFTs.

As a first step towards this classification, one can ask to describe how many
HWR \( {\mathcal{V}}_{c}(\Delta ) \) are there at a fixed \( c \). In general
any value of \( c \) and \( \Delta  \) could work, but if we require 
\emph{unitarity},
i.e. absence of negative norm states, then we can state the following 
results\cite{fqs}: 

\begin{enumerate}
\item for \( c<0 \) no representation is unitary 
\item for \( 0\leq c<1 \) the following set of \( {\mathcal{V}}_{c}(\Delta ) \)
is unitary \begin{equation}
\label{cvir}
c=1-\frac{6}{p(p+1)}\qquad ,\qquad p=2,3,4,...
\end{equation}
\begin{equation}
\label{deltavir}
\Delta =\Delta _{rs}=\frac{[(p+1)r-ps]^{2}-1}{4p(p+1)}\qquad 1\leq s\leq r\leq 
p-1\qquad r,s\in \mathbb {Z}
\end{equation}

\item for \( c\geq 1 \) all representations are unitary 
\end{enumerate}
Therefore a theory containing negative \( \Delta  \)'s or \( c \) is 
automatically
non-unitary. Unitarity is an essential requirement in string theory. Also many
statistical systems enjoy it, but there are well known cases (percolation, 
Lee-Yang
edge singularity) where unitarity does not hold (i.e. the Hamiltonian is not
real).

\subsection{The OPE-algebra and correlation functions}

All fields in a CFT are expected to obey a closed algebra under OPE 
\begin{equation}
A_{i}(z,\bar{z})A_{j}(0,0)=\sum _{k}B_{ij}^{k}(z,\bar{z})A_{k}(0,0)
\end{equation}
 where \( i,j,k \) run over all fields (primaries and secondaries) of the 
theory.
Conformal invariance (via eq.(\ref{Tphi})) reduces the problem to the knowledge
of the OPE-algebra among primary fields and puts constraints on the form of
the functions \( B_{ij}^{k}(z,\bar{z}) \). The OPE-algebra of primary fields
reads as \begin{equation}
\label{phiphi}
\phi _{i}(z,\bar{z})\phi _{j}(0,0)=\sum _{k}C_{ij}^{k}z^{\Delta _{k}-\Delta 
_{i}-\Delta _{j}}\bar{z}^{\bar{\Delta }_{k}-\bar{\Delta }_{i}-\bar{\Delta 
}_{j}}[\phi _{k}(0,0)]
\end{equation}
 where now the indices \( i,j,k \) run over all primaries of the theory and
\( [\phi _{k}(0,0)] \) means contribution from the whole conformal family \( 
[\phi _{k}] \),
which can be seen as an expansion over all secondaries of \( [\phi _{k}] \),
whose coefficients are also (in principle) fixed by conformal invariance. The
only objects that remain unfixed are the \emph{structure constants} \( 
C_{ij}^{k} \).
Were these known, one could reduce via iterative applications of OPEs 
(\ref{phiphi})
all the correlators among primaries to 2 and 3 point functions, which are fixed
by projective invariance \begin{equation}
\langle \phi _{1}(z_{1},\bar{z}_{1})\phi _{2}(z_{2},\bar{z}_{2})\rangle 
=\delta _{12}z_{12}^{2\Delta _{1}}\bar{z}_{12}^{2\bar{\Delta }_{i}}
\end{equation}
\begin{equation}
\langle \phi _{1}(z_{1},\bar{z}_{1})\phi _{2}(z_{2},\bar{z}_{2})\phi 
_{3}(z_{3},\bar{z}_{3})\rangle =C_{12}^{3}z_{12}^{\gamma _{12}}z_{13}^{\gamma 
_{13}}z_{23}^{\gamma _{23}}\bar{z}_{12}^{\bar{\gamma 
}_{12}}\bar{z}_{13}^{\bar{\gamma }_{13}}\bar{z}_{23}^{\bar{\gamma }_{23}}
\end{equation}
 where \( z_{ab}=z_{a}-z_{b} \), \( \gamma _{ab}=\Delta _{c}-\Delta 
_{a}-\Delta _{b} \)
with \( a\neq b\neq c \) and \( a,b,c=1,2,3 \). Correlators among secondaries
could be reduced to correlators among primaries via the OPE (\ref{Tphi}). So,
at least in principle, all correlators in the theory are computable: the theory
is exactly solvable. Unfortunately in most cases we do not know the constants
\( C_{ij}^{k} \). Constraints on \( C_{ij}^{k} \) come from the requirement
of associativity of the OPE-algebra, which is equivalent to ask \emph{duality}
of the 4-point functions. So in general \( C_{ij}^{k} \) can be computed if
we know the 4-point functions.

\subsection{Null vectors and minimal models}

It can happen that in a certain HWR a secondary state \( |\chi \rangle  \)
(to be called \emph{null vector}) satisfies \( L_{0}|\chi \rangle =(\Delta 
+N)|\chi \rangle  \)
and \( L_{n}|\chi \rangle =0 \), \( n>0 \). This means that \( |\chi \rangle  
\)
behaves like a primary state. This apparent contradiction is solved by saying
that the HWR is not irreducible. Taking \( |\chi \rangle \equiv 0 \) (that
can always be done consistently as it is easy to prove that \( \langle s|\chi 
\rangle =0 \)
for all \( |s\rangle \in  \) HWR, including \( |\chi \rangle  \) itself) we
get rid of the representation embedded in the HWR. If there are more than one
null vector, we repeat this procedure until we have {}``cleaned{}'' the 
representation
from all null vectors. What remains is a true irreducible module \( 
{\mathcal{V}}_{c}(\Delta ) \)
that can be used to build up a CFT.

When null vectors appear, they give constraints in the form of partial 
differential
equations on the correlators of primary fields. Indeed, consider a correlator
\( \langle \chi (z)\phi _{1}(z_{1})...\phi _{n}(z_{n})\rangle  \) where \( 
\phi _{i}(z_{i}) \)
are primary fields and \( \chi (z) \) is a null vector in the conformal family
\( [\phi ] \). Then \( \chi (z) \) is a secondary of \( \phi (z) \) and 
eq.(\ref{Tphi})
(or more precisely the conformal Ward identity obtained by it) implies the 
existence
of a suitable differential operator \( {\mathcal{D}} \) such that \( \langle 
\chi (z)\phi _{1}(z_{1})...\phi _{n}(z_{n})\rangle ={\mathcal{D}}\langle \phi 
(z)\phi _{1}(z_{1})...\phi _{n}(z_{n})\rangle  \).
On the other hand \( \chi \equiv 0 \), so that we have \begin{equation}
\label{diff}
{\mathcal{D}}\langle \phi (z)\phi _{1}(z_{1})...\phi _{n}(z_{n})\rangle =0
\end{equation}
 which is a differential equation to be satisfied by the correlators among 
primary
fields. This constraint can be so powerful to select a finite number of primary
fields under which the OPE-algebra closes. In this case only a finite set of
\( {\mathcal{V}}_{c}(\Delta ) \) are used to build up the model. We shall speak
in this case of a \emph{minimal} model of \( {\mathcal{V}} \). Minimal models
exist for \( c<1 \) only (there is a theorem by Cardy \cite{cardy86} that
prevents from the possibility to construct a model with finite number of \( 
{\mathcal{V}}_{c}(\Delta ) \)
for \( c\geq 1 \)). More precisely unitary minimal models can exist only for
the values of \( c=\frac{1}{2},\frac{7}{10},\frac{4}{5},\frac{6}{7},... \)
given by the formula (\ref{cvir}) and they can be built up using only the \( 
{\mathcal{V}}_{c}(\Delta ) \)
representations such that \( \Delta  \) is contained in the \emph{Kac-table}
given by eq.(\ref{deltavir}).

The differential equations (\ref{diff}) reduce, in the case of 4-point 
functions
and after mapping \( z_{1}\rightarrow 0,z_{2}\rightarrow 1,z_{4}\rightarrow 
\infty  \),
to ordinary fuchsian differential equations in \( z\equiv z_{3} \). In the
case of minimal models a solution to these equations can be given 
\cite{dotfat}.
This means that the \( C_{ij}^{k} \) can be computed, and so exact solvability
of the minimal models is ensured.

\subsection{CFT on a cylindrical geometry and FSE}

The mapping of the complex plane into a strip of thickness \( L \) with 
periodic
boundary conditions is a conformal transformation, realized by the 
mapping\begin{equation}
\label{plane-cyl}
z=e^{\frac{2\pi }{L}u}\quad ,\quad \bar{z}=e^{\frac{2\pi }{L}\bar{u}}
\end{equation}
with \( u=\tau +i\sigma  \) and \( \bar{u}=\tau -i\sigma  \), where \( \tau 
\in ]-\infty ,+\infty [ \)
is the cylinder {}``time{}'' and \( \sigma \in [0,L[ \) the cylinder 
{}``space{}''.
Under conformal mappings, the stress energy tensor transforms as\[
T(z)=\left( \frac{\partial u}{\partial z}\right) ^{2}T(u)+\frac{c}{12}\left[ 
\frac{\partial ^{3}_{z}u}{\partial _{z}u}-\frac{3}{2}\left( \frac{\partial 
_{z}^{2}u}{\partial _{z}u}\right) ^{2}\right] \]
 Consequently, the modes of \( T \) on the plane map on those on the cylinder
according to\[
T(u)=-\frac{c}{24}+\sum _{n\in \mathbb Z}L_{n}e^{inu}\]
i.e. they become Fourier modes on the cylinder. The radial quantization in \( 
z \)
briefly mentioned above becomes in this cylindrical context a usual time 
ordering
quantization. Periodic boundary conditions mean \( T(u+L)=T(u) \). The physical
interpretation of modes on the cylinder is that total energy and momentum of
the system are given by the zero modes\[
E=\frac{2\pi }{L}(L_{0}+\bar{L}_{0}-\frac{c}{12})\quad ,\quad P=\frac{2\pi 
}{L}(L_{0}-\bar{L}_{0})\]
 Therefore a state \( |\Delta ,\bar{\Delta }\rangle  \) on the cylinder 
represents
a state of given energy and momentum\[
E=\frac{2\pi }{L}(\Delta +\bar{\Delta }-\frac{c}{12})\quad ,\quad P=\frac{2\pi 
}{L}(\Delta -\bar{\Delta })\]
In particular the vacuum of a unitary theory (\( \Delta =\bar{\Delta }=0 \))
has energy\[
E_{vac}=-\frac{\pi c}{6L}\]
 given by the Casimir effect of being on a finite and periodic geometry. We
see that the constant \( c \) has the physical meaning of measuring this 
Casimir
effect.

\subsection{Integrability of CFT\label{integrableCFT}}

CFT is an integrable QFT. This can be shown by explicitly constructing an 
infinite
set of conserved charges. Define the charges\[
I_{s+1}=\frac{1}{L}\int _{0}^{L}d\sigma \, T_{s}(u)=\oint dz\, T_{s}(z)\]
 where the currents \( T_{s}(u) \) are time ordered (in \( \tau  \) time)
polynomials in \( T \) and its derivatives at level \( s \) in the Verma module
of the identity operator. One can show that if \( s \) is odd, all the \( 
T_{s} \)
fields at level \( s \) can be arranged to be total derivatives of some \( 
T_{s-1} \)
and therefore their integrals \( I_{s+1} \) are trivially zero. Nontrivial
integrals of motion generated by \( T \) exist only for even \( s \). At a
given level there can be more than one monomial in \( T \) and its derivatives,
the most general current will be a linear combination of these. If we require
that the corresponding charges commute with all the ones at smaller values of
\( s \) then the coefficients turn out to be fixed in a unique way, thus giving
a sequence of currents, one for each even level \( s \), that are in 
involution,
i.e mutually commute. This actually defines an integrable QFT. Here are the
first few currents of this kind.\[
\begin{array}{c}
T_{2}(u)=T(u)\\
T_{4}(u)=:T(u)^{2}:\\
T_{6}(u)=:T(u)^{3}:+\frac{c+2}{12}:(\partial _{u}T(u))^{2}:
\end{array}\]
etc... Of course a similar sequence of integrals \( \bar{I}_{s+1} \) holds
also for \( \bar{T} \). Note that the first current is \( T \) itself. 
Therefore,
energy and momentum are conserved quantities belonging to this set, and build
up a vector of spin 1. All the other charges, as they commute with the 
Hamiltonian,
are also conserved charges in an infinite sequence of higher and higher odd
spins.

Of course, if the model possesses higher symmetries (like in \( W \)-algebras
or in WZW models) there can be other sets of conserved currents and the 
sequence
of spin of the conserved charges could be different.

\subsection{c=1 Conformal Field Theory\label{section:free_boson}}

A particular example of CFT that will be used frequently in the sequel of this
review is the case of the so called \emph{massless free boson} compactified
on a target space of a circle of radius \( R \). Here we give a brief summary
of such a \( c=1 \) CFT. The Lagrangian is taken to be

\begin{equation}
\label{bosone_{l}ibero}
\displaystyle {\cal L}=\frac{1}{8\pi }\int ^{L}_{0}\partial _{\mu }\varphi 
(\sigma ,\, \tau )\partial ^{\mu }\varphi (\sigma ,\, \tau )d\sigma \: ,\: 
\sigma \in [0,L]\: ,
\end{equation}
 where \( L \) is the spatial volume (i.e. the theory is defined on a 
cylindrical
space-time with circumference \( L \)). The superselection sectors are 
classified
by the \( \widehat{U(1)}_{L}\times \widehat{U(1)}_{R} \) Kac-Moody symmetry
algebra, generated by the currents

\[
J(z)=i\partial _{z}\varphi \, ,\quad \bar{J}(\bar{z})=i\partial 
_{\bar{z}}\varphi \, .\]
 where we have mapped the cylinder back to the plane according to 
eq.(\ref{plane-cyl}).
The left/right moving energy-momentum tensor is given by \[
T(z)=\frac{1}{8\pi }\partial _{z}\varphi \partial _{z}\varphi =\sum ^{\infty 
}_{k=-\infty }L_{k}z^{-k-2}\quad ,\quad \bar{T}(\bar{z})=\frac{1}{8\pi 
}\partial _{\bar{z}}\varphi \partial _{\bar{z}}\varphi =\sum ^{\infty 
}_{k=-\infty }\bar{L}_{k}\bar{z}^{-k-2}\]
 The coefficients \( L_{n} \) and \( \bar{L}_{n} \) of the Laurent expansion
of these fields generate two mutually commuting Virasoro algebras. If the 
quasi-periodic
boundary conditions are required for the boson

\[
\varphi (\sigma +L,\tau )=\varphi (\sigma ,\tau )+2\pi mR\, ,\quad m\in 
\mathbb {Z}\, ,\]
 then the sectors are labelled by a pair of numbers \( (n,\, m) \), where \( 
\frac{n}{R} \)
(\( n \) is half integer because of the locality, see later) is the eigenvalue
of the total field momentum \( \pi _{0} \)

\[
\pi _{0}=\int ^{L}_{0}\pi (\sigma ,\tau )d\sigma \, ,\quad \pi (\sigma ,\tau 
)=\frac{1}{4\pi }\partial _{\tau }\varphi (\sigma ,\tau )\: ,\]
 and \( m \) is the winding number, i.e. the eigenvalue of the topological
charge \( Q \) defined by

\[
Q=\frac{1}{2\pi R}\int ^{L}_{0}\partial _{\sigma }\varphi (\sigma ,\tau 
)d\sigma \, .\]
 In the sector with quantum numbers \( (n,\, m) \), the scalar field is 
expanded
in modes as follows: \[
\begin{array}{rl}
\displaystyle \varphi (\sigma ,\tau )= & \phi (z)+\bar{\phi }(\bar{z})\: ,\\
\phi (z)= & \frac{1}{2}\varphi _{0}-ip_{+}\log z+i\sum _{k\neq 
0}\frac{1}{k}a_{k}z^{-k}\: ,\\
\bar{\phi }(\bar{z})= & \frac{1}{2}\varphi _{0}-ip_{-}\log \bar{z}+i\sum 
_{k\neq 0}\frac{1}{k}\bar{a}_{k}\bar{z}^{-k}\: ,
\end{array}\]
 where the left and right moving field momenta \( p_{\pm } \) (which are in
fact the two \( U(1) \) Kac-Moody charges) are given by

\begin{equation}
\label{p+-}
\displaystyle p_{\pm }=\frac{n}{R}\pm \frac{1}{2}mR\, .
\end{equation}
 The Virasoro generators take the form

\[
L_{n}=\frac{1}{2}\sum ^{\infty }_{k=-\infty }:a_{n-k}a_{k}:\, ,\quad 
\bar{L}_{n}=\frac{1}{2}\sum ^{\infty }_{k=-\infty 
}:\bar{a}_{n-k}\bar{a}_{k}:\, ,\]
 where the colons denote the usual normal ordering, according to which the 
oscillator
with the larger index is put to the right.

The ground states of the different sectors \( (n,\, m) \) are created from
the vacuum by the (Kac-Moody) primary fields, which are vertex operators of
the form

\begin{equation}
\label{vertex_{o}perators}
V_{(n,\, m)}(z,\overline{z})=:\exp i(p_{+}\phi (z)+p_{-}\bar{\phi 
}(\overline{z})):\: .
\end{equation}
 The left and right conformal weights of the field \( V_{(n,\, m)} \) (i.e.
the eigenvalues of \( L_{0} \) and \( \bar{L}_{0} \)) are given by the formulae

\begin{equation}
\label{delta+-_{v}ertex_{o}p}
\displaystyle \Delta ^{\pm }=\frac{p^{2}_{\pm }}{2}.
\end{equation}
 The Hilbert space of the theory is given by the direct sum of the Fock modules
built over the states

\begin{equation}
\label{fock_{s}tates}
\left| n,\, m\right\rangle =V_{(n,\, m)}(0,0)\left| vac\right\rangle \: ,
\end{equation}
 with the help of the creation operators \( a_{-k}\: ,\: \bar{a}_{-k}\: k>0 \):

\[
{\mathcal{H}}=\bigoplus _{(n,\, m)}\{a_{-k_{1}}\ldots 
a_{-k_{p}}\bar{a}_{-l_{1}}\ldots \bar{a}_{-l_{q}}|n,\, m\rangle ,\, 
k_{1},\ldots \, k_{p},\, l_{1},\ldots \, l_{q}\in \mathbb {Z}_{+}\}\]
 The boson Hamiltonian on the cylinder is expressed in terms of the Virasoro
operators as

\begin{equation}
\label{cylinder_{h}amilt}
\displaystyle H_{CFT}=\frac{2\pi }{L}\left( 
L_{0}+\overline{L}_{0}-\frac{c}{12}\right) \: ,
\end{equation}
 where the central charge is \( c=1 \). The generator of spatial translations
is given by

\begin{equation}
\label{cylinder_{m}oment}
\displaystyle P=\frac{2\pi }{L}\left( L_{0}-\bar{L}_{0}\right) \; .
\end{equation}
 The operator \( L_{0}-\bar{L}_{0} \) is the conformal spin which has 
eigenvalue
\( nm \) on the primary field \( V_{(n,\: m)} \).

One can also introduce twisted sectors using the operator \( {\mathcal{T}} \)
that performs spatial translations by \( L \): \( x\rightarrow x+L \). The
primary fields \( V_{(n,\, m)} \) as defined above satisfy the periodicity
condition \( {\mathcal{T}}V_{(n,\, m)}=V_{(n,\, m)}. \) If the more general
twisted boundary condition labelled by a real parameter \( \nu  \) is required
\[
{\mathcal{T}}V_{(n,\, m)}=\exp \left( i\nu Q\right) V_{(n,\, m)}\, ,\]
 then it is possible to generate superselection sectors for which \( n\in 
\mathbb {Z}+\frac{\nu }{2\pi } \).

It is important to stress that a particular \( c=1 \) CFT is specified by 
giving
the spectrum of the quantum numbers \( (n,m) \) (and the compactification 
radius
\( R \)) such that the corresponding set of vertex operators (and their 
descendants)
forms a \emph{closed and local} operator algebra. The locality requirement is
equivalent to the fact that the operator product expansions of any two such
local operators is single valued in the complex plane of \( z \). This 
condition,
which is weaker than the modular invariance of the CFT, is the adequate one
since the theory is considered on a space-time cylinder and we do not wish to
define it on higher genus surfaces.

By this requirement of locality, it was proved in \cite{kl-me} that there are
only two maximal local subalgebras of vertex operators: \( {\mathcal{A}}_{b} \)
generated by the vertex operators \[
\{V_{(n,\, m)}:\, n,\, m\in \mathbb {Z}\}\, ,\]
 and \( {\mathcal{A}}_{f} \) generated by \[
\{V_{(n,\, m)}:\, n\in \mathbb {Z},\, m\in 2\mathbb {Z}\: or\: n\in \mathbb 
{Z}+\frac{1}{2},\, m\in 2\mathbb {Z}+1\}\, .\]
 Other sets of vertex operators can be built, but the product of two of them
gives a nonlocal expression.

\section{Perturbed Conformal Field Theory}

In the previous sections we have seen that a QFT can be thought as a RG flow
in the space of actions, and its UV and IR points correspond to scale invariant
CFT. We have then concentrated on CFT and explored its properties. Armed with
this new insight we may wonder if the CFT data are able to predict some piece
of information about the theory out of the fixed points.

We may think to define a QFT as a deformation of a critical CFT by some 
operators
\cite{Zam-adv}, i.e. to assume as action the following 
expression\begin{equation}
\label{pcft-action}
S=S_{CFT}+\sum _{i=1}^{n}\lambda _{i}\int d^{2}x\Phi _{i}(x)
\end{equation}
 Of course, the class of QFT\( _{2} \) is larger than the one described by
this sort of action. Nevertheless, this class of the so called \emph{Perturbed
Conformal Field Theories} (PCFT) is very important because most of the 
predictions
can be done working in this realm and many applications to the vicinity of 
critical
points in the theory of critical phenomena can be described by this class of
actions. 

Lorentz invariance in 1+1 dimensions is equivalent to rotational invariance
in the 2 dimensional space once a Wick rotation is performed. Hence, to achieve
Lorentz invariance we must require that the fields \( \Phi _{i} \) are scalars
under 2 dimensional rotations, i.e. their conformal spin \( \Delta 
_{i}-\bar{\Delta }_{i} \)
must be null, which implies \( \Delta _{i}=\bar{\Delta }_{i} \). Their 
anomalous
dimensions are therefore \( 2\Delta _{i} \).

\subsection{Relevant, Irrelevant and Marginal Perturbations}

Let us consider the simple case of one single perturbing field \( \Phi  \),
of left conformal dimension \( \Delta  \). The coupling \( \lambda  \) is
therefore dimensionful, with scaling dimension \( y=2-2\Delta  \). 
Renormalization
is needed \cite{Cappelli-Latorre}, and can be achieved by introducing a 
dimensionless
coupling \( g \), a mass scale \( \mu  \) and a renormalized field \( \Phi 
(x,g) \)
by the relations\[
g=\mu ^{-y}Z_{g}\lambda \quad ,\quad \Phi (x)=\sqrt{Z_{\Phi }}\Phi (x,g)\]
 where \( Z_{g} \) and \( Z_{\Phi } \) are renormalization prefactors 
collecting
all infinite contributions.

\( g=0 \) is a fixed point, because the interaction perturbing term disappears
and one is left with the pure \( S_{CFT} \). The field \( \Phi =\frac{\partial 
L}{\partial g} \)
must have a limit in the unperturbed CFT, say \( \Phi _{0} \), with 
\begin{equation}
\label{gamma0}
\Gamma (0)\Phi _{0}=\gamma (0)\Phi _{0}=2\Delta \Phi _{0}
\end{equation}
We are asking which are the conditions on the field \( \Phi  \) that guarantee
that the fixed point at \( g=0 \) is an UV one, from which the RG flow actually
gets out and, increasing \( t \) goes to some IR destiny. To have this, \( g \)
must \emph{increase} towards an IR fixed point \( g^{*} \), and the anomalous
dimension of \( \Phi  \) changes according to the already seen 
formula\begin{equation}
\label{gamma}
\Gamma (g)\Phi =\left( 2-\frac{\partial \beta }{\partial g}\right) \Phi 
\end{equation}
 Putting together eqs.(\ref{gamma0}) and (\ref{gamma}) we can compute the
first order of the beta function: \( \beta (g)=yg+O(g^{2}) \). The nature of
the fixed point (UV or IR) is chosen by the sign of the first derivative of
\( \beta (g) \), as we saw in section 2. \( y=\beta '(0)>0 \) only if \( 
\Delta <1 \). 

Fields are classified with respect to the RG group as

\begin{itemize}
\item \emph{relevant} if \( \Delta <1 \). If such a field perturbs a conformal 
action,
it creates exactly the situation described above, i.e. the theory starts to
flow along a RG trajectory going to some IR destiny.
\item \emph{irrelevant} if \( \Delta >1 \). Such fields correspond to non 
renormalizable
perturbations which describe the neighborhood of IR non trivial fixed points.
Usually one speaks of an \emph{attraction} field better than a perturbing one.
We shall not deal with this case in the following, but the interested reader
may consult, for example \cite{Fev-Quattr-Rav} to see some possible 
applications
of this situation.
\item \emph{marginal} if \( \Delta =1 \). These are further distinguished by 
the
behaviour of higher derivatives of the beta function

\begin{itemize}
\item if they stay marginal at all orders in perturbation theory, they are 
called
\emph{truly marginal}, and describe deformation of a CFT leading to another
CFT with the same value of the central charge. They identify a submanifold of
CFT in the space of actions.
\item if instead the anomalous dimension becomes smaller than 2 at higher 
orders in
perturbation theory, the fields are called \emph{marginally relevant}. They
provide a perturbation of CFT defining a non scale invariant theory, but the
perturbation theory is much more complicated and in particular the smoothness
of the deformation of the Hilbert Space is no more valid. This sort of UV 
limit,
that Lüscher defined as \emph{singular} many years ago, is typical of the 
asymptotically
free theories and hence very interesting for the properties that such theories
share with QCD.
\item finally there can be some marginally irrelevant fields, defining a non 
renormalizable
attraction to an IR no trivial point of particular theories
\end{itemize}
\end{itemize}

\subsection{Conformal Perturbation Theory}

In a PCFT it is natural to define a Conformal Perturbation Theory (CPT). This
is a perturbation theory where the unperturbed part of the action is not taken,
as usual, to be the free part, but better the CFT at UV. A general correlator
\( \langle X\rangle  \) of a string of fields \( X=A_{1}...A_{N} \) expresses
as\[
\langle X\rangle =\int {\cal D}\varphi \, Xe^{-S[\varphi ]}=\int {\cal 
D}\varphi \, Xe^{-S_{CFT}-\lambda \int d^{2}x\Phi (x)}\]
and expanding in powers of \( \lambda  \) one can express \( \langle X\rangle  
\)
as a series of conformal correlators (in principle computable by CFT 
techniques)
with the insertion of the perturbing field\[
\langle X\rangle =\sum _{k=0}^{\infty }\frac{(-\lambda )^{k}}{k!}\int \langle 
X\Phi (x_{1})...\Phi (x_{k})\rangle _{CFT}d^{2}x_{1}...d^{2}x_{k}\]
The possible IR divergences of such an expansion are cured if we put the theory
on a finite geometry, e.g. on a periodic cylinder as above. A full treatment
and detailed expressions for the coefficients of the CPT on the cylinder are
given in \cite{kl-me-cpt}. CPT turns out to be UV finite if \( y>1 \) i.e.
\( \Delta <\frac{1}{2} \). For \( 0<y\leq 1 \) only a \emph{finite} number
of terms are divergent: the action (\ref{pcft-action}) is 
\emph{super-renormalizable}
in this case. Once the divergences cured (usually analytic continuation in \( 
y \)
does the job cleanly) the series are convergent within a finite radius (which
is not the case in general for perturbation series of higher dimensional QFT).
It can even happen that the IR point \( g^{*} \) falls into the convergence
region, like it has been shown in \cite{Zam86,Ludwig-Cardy} for the minimal
models perturbed by \( \Phi _{1,3} \) at large \( p \). In this case, one
can do predictions on the RG flow and follow scaling functions of physical 
quantities
by use of the CPT. In general, the radius of convergence is too small to 
contain
the IR region of the RG flow. Nonperturbative effects take place and one has
to resort to alternative methods of calculation to study scaling functions in
the IR regime in order to relate them to the particle content of the theory.

For \( y=0 \) one exactly gets the case of an asymptotically free QFT. The
perturbing operator is marginally relevant and the theory is renormalizable
but not super-renormalizable, i.e. there are in general infinities at each 
order,
but they are cured by the insertion of a finite number of counterterms. 
Finally,
for \( y<0 \) the theory is non-\emph{renormalizable} and in general there
are new infinities and new counterterms to be added at any order, thus loosing
predictability. However, in the case of \emph{integrable} theories, one can
go a bit further \cite{Fev-Quattr-Rav} in restricting the form of counterterms
and getting some physical information on the attracting field (these theories
better describe the attraction to an IR point of a RG flow by some irrelevant
operator).

\subsection{Integrability in PCFT}

We have seen that any CFT is an integrable theory and possesses an infinite
set of mutually commuting charges. The perturbation (\ref{pcft-action}) usually
destroys such conservation laws. It can happen, however, that a subset of the
local charges of section \ref{integrableCFT} keeps conserved because the 
modification
of the current due to the perturbation can still be arranged as total 
derivative,
thus recovering the continuity equation even off-criticality. In more precise
words, the equation \( \partial _{\bar{z}}T_{s}(z)=0 \) valid in the 
unperturbed
CFT is no longer valid off-criticality. It is deformed by contributions coming
from the perturbation series\begin{equation}
\label{pert-current}
\partial _{\bar{z}}T_{s}=\lambda R^{(1)}_{s-1}+\lambda 
^{2}R_{s-1}^{(2)}+...+\lambda ^{N}R^{(N)}_{s-1}
\end{equation}
 The fields \( R_{s-1}^{(n)} \) must have, in the super-renormalizable case,
a limit in the unperturbed CFT, defined by some field in the algebra of fields
of the CFT, with conformal dimensions \( (\Delta ^{(n)},\bar{\Delta }^{(n)}) \)
appearing in the Kac table of primaries or secondaries. Dimensional balancing
of eq.(\ref{pert-current}) gives\begin{equation}
\label{pert-dimensions}
\Delta ^{(n)}=s-n(1-\Delta )\qquad \bar{\Delta }^{(n)}=1-n(1-\Delta )
\end{equation}
 where \( \Delta  \) is the dimension of the perturbing field \( \Phi  \).
This implies that the series (\ref{pert-current}) actually truncates at some
integer \( N \), as conformal dimensions in a CFT are bounded by below. In
general there is only one field in the Kac table compatible with 
(\ref{pert-dimensions})
appearing in (\ref{pert-current}). In this case we speak of a no-resonance
perturbation. Concentrating on this simple case, where only \( R_{s-1}^{(1)} \)
appears with conformal dimensions \( (s-1+\Delta ,\Delta ) \), we see that
such field must be a secondary of the perturbing field \( \Phi  \) of spin
\( s-1 \). It is obtained by applying only left modes \( L_{-n} \) to \( \Phi  
\).
If the combination of modes can be put in the form of\[
L_{-1}\cdot \mbox {(any\, combination\, of\, strings\, of\, }L_{-n}\mbox 
{'s)}\]
 then, as \( L_{-1}\equiv \partial _{z} \), one can see that \( \lambda 
R_{s-1}^{(1)}=\partial _{z}\Theta _{s-2} \)
for some field \( \Theta _{s-2} \) also in the secondaries of \( \Phi  \)
but one level below. Then eq.(\ref{pert-current}) boils down to a genuine 
continuity
equation \[
\partial _{\bar{z}}T_{s}=\partial _{z}\Theta _{s-2}\]
guaranteeing a conservation of the charge\[
I_{s+1}=\oint dz\, T_{s}(z,\bar{z})+\oint d\bar{z}\, \Theta _{s-2}(z,\bar{z})\]

If such a situation is realizable or not depends on the nature of the 
perturbation
\( \Phi  \). A. Zamolodchikov \cite{Zam-adv} has given a sufficient but not
necessary condition to ensure the presence of a conserved current, based on
the analysis of the dimensions of subspaces of the Verma modules of \( \Phi  \)
and the identity. For details see the original article \cite{Zam-adv}.

Integrability of a two-dimensional massive QFT has strong implications on the
properties of its S-matrix \cite{zam79}.

\begin{enumerate}
\item the number of particles is conserved in the scattering process, there is 
no
particle production. Only exchanges of internal quantum numbers are allowed
among particles in a multiplet;
\item the final set of momenta coincides with the initial one: \( 
\{p_{1},....,p_{n}\}_{in}\equiv \{p'_{1},....,p'_{n}\}_{out} \);
\item the S-matrix for a process with \( n \) incoming and \( n \) outgoing 
particles
factorizes into S-matrices of 2 incoming and 2 outgoing particles. To guarantee
that the order of this factorization is irrelevant, the 2 particle S-matrices
have to satisfy a factorization equation known as \emph{Yang-Baxter} equation
(here 1,2,3 label the particles and \( \theta _{ij}=\theta _{i}-\theta _{j} \),
\( \theta _{i} \) being the rapidity of particle \( i \))\[
S_{i_{1}i_{2}}^{k_{1}k_{2}}(\theta _{12})S_{k_{1}i_{3}}^{j_{1}k_{3}}(\theta 
_{13})S_{k_{2}k_{3}}^{j_{2}j_{3}}(\theta 
_{23})=S_{k_{1}k_{2}}^{j_{1}j_{2}}(\theta 
_{12})S_{i_{1}k_{3}}^{k_{1}j_{3}}(\theta 
_{13})S_{i_{2}i_{3}}^{k_{2}k_{3}}(\theta _{23})\]

\end{enumerate}
The factorizable S-matrix of an integrable PCFT can often be conjectured by
the knowledge of integrals of motion, that constrain the bootstrap equations
and by the Yang-Baxter Equation itself, that imposes severe constraints on the
matricial form of \( S_{ab}^{cd}(\theta ) \). However, this conjecture has
to be verified against independent checks. The particle description of the 
Hilbert
space must be linked with the conformal description valid in the vicinity of
the critical point, otherwise one cannot assert that the two theories defined
as perturbation of a CFT and as factorized scattering theory are the same.

\subsection{Scaling functions}

One possible bridge between the two formulations -- perturbed CFT versus 
Factorizable
Scattering Theory -- is given by FSE, more specifically by the use of the so
called \emph{scaling} \emph{functions.} We have seen that the energy of a state
\( |i\rangle  \) on a cylinder of circumference \( L \) in CFT is given 
by\begin{equation}
\label{cardy}
E_{i}=-\frac{\pi c_{i}}{6L}\quad \mbox {with}\quad c_{i}=c-12(\Delta 
_{i}+\bar{\Delta }_{i})
\end{equation}
 It is obvious that when we move outside the critical point such energy level
has a different dependence on \( L \). However, the overall \( 1/L \) 
dependence
is fixed by dimensional reasons, so all the variation of the dependence must
be confined in a dimensionless function that we are free to normalize in such
a way that in the limit \( L\rightarrow 0 \) it reproduces eq.(\ref{cardy}).
Being dimensionless, such a function can depend on \( L \) only in a 
dimensionless
way, i.e. we are forced to introduce a mass parameter \( m \) to compensate
for the physical dimension of \( L \). The dependence will be on a 
dimensionless
parameter \( l=mL \). This is very reminiscent of what already illustrated
in the introduction and the discussions made there about the UV and IR limits
recovered for \( l\rightarrow 0 \) and \( l\rightarrow \infty  \) respectively
of course apply here as well. They show that for each state \( |i\rangle  \)
in the theory there is a function \( c_{i}(l) \) attached to this state in
such a way that its energy is\[
E_{i}(L)=-\frac{\pi c_{i}(l)}{6L}\]
such that

\begin{itemize}
\item \( \lim _{l\rightarrow 0}c_{i}(l)=c_{i}=c-12(\Delta _{i}+\bar{\Delta 
}_{i}) \)
i.e. it reconstruct conformal data in the limit \( l\rightarrow 0 \);
\item it can help, in a manner that will become apparent later, to extract 
scattering
data in the limit \( l\rightarrow \infty  \)
\end{itemize}
The forced introduction of a mass scale parameter should not surprise: we are
actually going out of a critical CFT invariant point into a theory which is
not scale invariant. It is therefore natural that a mass scale should be 
introduced.

To such functions \( c_{i}(l) \) the name of \emph{scaling} \emph{functions}
is usually given. The problem to bridge between the PCFT and the Factorized
Scattering formulations of an integrable QFT\( _{2} \) is recast into the 
computation
of such scaling functions.

Of course, one way to compute them, also in non-integrable theories, is to 
resort
to CPT. However, as already commented, the validity of CPT is normally confined
to a small radius of convergence that does not allow to make contact with the
IR region where scattering theory is apparent. This justifies the seek for the
development of nonperturbative methods.

\begin{itemize}
\item \textcolor{black}{The} \textcolor{black}{\emph{Truncated Conformal Space 
Approach}}
\textcolor{black}{(TCSA) \cite{yurov-zam} consists in diagonalizing the 
truncated
(\( E<E_{cut} \)) Hamiltonian \( H_{CFT}+V_{pert} \) numerically. It is 
non-perturbative,
and applicable even to non-integrable QFT. However, it is affected by 
truncation
errors} \textbf{\textcolor{black}{}}\textcolor{black}{that increase with \( l 
\)
and, being totally numeric, does not allow for any} 
\textbf{\textcolor{black}{}}\textcolor{black}{analytic
control of the scaling functions. }
\item \textcolor{black}{The} \textcolor{black}{\emph{Thermodynamic Bethe 
Ansatz}}
\textcolor{black}{(TBA) \cite{zam_TBA} implements the thermodynamics of a gas
of particles governed by the given S-matrix}\textbf{\textcolor{black}{.}} 
\textcolor{black}{The
free energy can be exactly computed in terms of a set of coupled non-linear
integral equations. Exchanging the role of time and space, such free energy
can be reinterpreted as a FSE Casimir vacuum energy where the role of \( L \)
is played by the temperature. Recently also excited states have been accessed
with this method \cite{dorey-tateo}, that has been very popular for the last
ten years. The treatment of the vast literature on this method is out of the
scope of this review. }
\item \textcolor{black}{In the next sections we shall illustrate an 
alternative method
to obtain scaling functions for an integrable model starting directly from its
definition as a QFT regularized on a lattice. This method is know in QFT as}
\textcolor{black}{\emph{Destri De Vega Non-Linear Integral Equation}} 
\textcolor{black}{(NLIE)
from light cone Bethe Ansatz \cite{ddv 95}, but similar equations have also
been derived in Condensed Matter Theory \cite{klumper}.}
\end{itemize}

\subsection{Sine-Gordon / massive Thirring models as 
PCFT\label{section:sine-gordon}}

The sine-Gordon model, which has been well known at the classical level for
the late fifty years, plays also an important role in quantum theory, thanks
to its particular properties of non-linearity and integrability. It has been
successfully applied to very different sectors of Mathematics and Physics, 
ranging
from partial differential equation theory to particle physics or solid state
physics. Recent applications of the classical model are related to nonlinear
optics (resonant dielectric media) and optical fibers, magnetic properties of
polymers, propagation of waves in crystals, etc... Interesting applications
of the quantum model are related to Kondo effect and to the thermodynamics of
some chemical compound, as Copper-Benzoate Cu(C\( _{6} \)D\( _{5} \)COO)\( 
_{2}\cdot  \)3DO\( _{2} \)
\cite{copper_benzoate}. At the same time, the quantum theory shows a 
phenomenology
that is similar to the Skyrme model used before QCD era to describe barions
and strong interactions.

The most relevant properties of the model are

\begin{itemize}
\item at a classical level, all the solutions of the equations of motion are 
known
(exact integrability via inverse scattering method) 
\item the classical solutions describe solitons, antisolitons and bound states 
(breathers);
in a scattering process this solutions are transparent (it is the mathematical
meaning of {}``soliton{}'') 
\item it admits, both at a classical and at the quantum level, a countable 
infinite
set of conserved charges 
\item the quantization of the theory describes an interacting particle with 
its antiparticle
and, in a certain (attractive) regime, bound states 
\item the S matrix has been exactly determined; only elastic scattering 
processes
can take place (i.e. no particle production), that is the quantum analog of
the classical transparency of solitons 
\end{itemize}
The minkowskian Lagrangian of sine-Gordon theory is given by

\begin{equation}
\label{sG_Lagrangian}
\displaystyle {\cal \mathcal{L}}_{sG}=\int _{-\infty }^{+\infty }\left( 
\frac{1}{2}\partial _{\nu }\Phi \partial ^{\nu }\Phi +\frac{\mu ^{2}}{\beta 
^{2}}:\cos \left( \beta \Phi \right) :\right) dx\, ,
\end{equation}
 where \( \Phi  \) denotes a real scalar field, while that of the massive 
Thirring
theory is of the following form:

\begin{equation}
\label{mTh_Lagrangian}
\displaystyle {\cal \mathcal{L}}_{mTh}=\int _{-\infty }^{+\infty }\left( 
\bar{\Psi }(i\gamma _{\nu }\partial ^{\nu }+m_{0})\Psi -\frac{g}{2}\bar{\Psi 
}\gamma ^{\nu }\Psi \bar{\Psi }\gamma _{\nu }\Psi \right) dx\, ,
\end{equation}
 describing a current-current selfinteraction of a Dirac fermion \( \Psi  \).
It is known that the two theories are deeply related provided their coupling
constants satisfy \[
\frac{\beta ^{2}}{4\pi }=\frac{1}{1+g/\pi }\, .\]
 For comparison with the Destri-de Vega nonlinear integral equation, it is 
important
to deal with cylindrical sine-Gordon and massive Thirring, i.e. the integrals
in (\ref{sG_Lagrangian}, \ref{mTh_Lagrangian}) must be taken in the interval
\( [0,\, L] \).

The \( \cos  \) term in (\ref{sG_Lagrangian}) can be considered as a 
perturbation
of the \( c=1 \) free boson compactified on a cylinder, as described in section
\ref{section:free_boson}. Similarly the massless (\( m_{0}=0 \)) Thirring
model is a \( c=1 \) conformal field theory and the mass term plays the role
of a perturbation. From Coleman's paper \cite{s-coleman} it is known that 
correlation
functions of the perturbing fields \( \bar{\Psi }\Psi  \) and \( :\cos \beta 
\Phi : \)
are identical, then both models can be considered as the perturbations of a
\( c=1 \) compactified boson by a potential \( V \) :

\begin{equation}
\label{hamilt_{p}erturb}
\displaystyle S_{sG/mTh}=S_{CFT}+V\quad ,\quad V=\lambda \int \left( 
V_{(1,0)}(z,\overline{z})+V_{(-1,0)}(z,\overline{z})\right) d^{2}x\: ,
\end{equation}
 which is related to the bosonic Lagrangian (\ref{sG_Lagrangian}) by the 
following
redefinitions of the field and the parameters:

\begin{equation}
\label{rinomina}
\displaystyle \varphi =\sqrt{4\pi }\Phi \: ,\quad R=\frac{\sqrt{4\pi }}{\beta 
}\: ,\quad \lambda =\frac{\mu ^{2}}{2\beta ^{2}}\: .
\end{equation}
 For later convenience, a new parameter \( p \) can be defined by

\begin{equation}
\label{parametro_p}
\displaystyle p=\frac{\beta ^{2}}{8\pi -\beta ^{2}}=\frac{1}{2R^{2}-1}\: .
\end{equation}
 The point \( p=1 \) (i.e. \( g=0 \)) is the free fermion point, corresponding
to a massive Dirac free fermion. The particle spectrum of Sine-Gordon for \( 
p>1 \)
is composed by the soliton (\( s \)) and its antiparticle, the antisoliton
(\( \bar{s} \)). It is known as repulsive regime because no bound states can
take place. \( p<1 \) is the attractive regime, because \( s \) and \( \bar{s} 
\)
can form bound states that are known as breathers. The values \( \displaystyle 
p=\frac{1}{k}\; ,\; k=1,2,\ldots  \)
are the thresholds where a new bound state appears. The potential term becomes
marginal when \( \beta ^{2}=8\pi  \) which corresponds to \( p=\infty  \).
The perturbation conserves the topological charge \( Q \), which can be 
identified
with the usual topological charge of the Sine-Gordon theory and with the 
fermion
number of the mTh model.

Mandelstam \cite{mandelstam} showed that a fermion operator satisfying the
massive Thirring equation of motion can be constructed as a nonlocal functional
of a pseudoscalar field (boson) satisfying the sine-Gordon equation. But the
fermion and the boson are not relatively local and then do not create the same
particle (the two theories are not equivalent).

The difference between them is that they correspond to the perturbation by the
same operator of the two \emph{different local} c=1 CFTs \( {\mathcal{A}}_{b} 
\)
and \( {\mathcal{A}}_{f} \) as in \ref{section:free_boson}. The short distance
behaviour of the sG theory is described by the local operator algebra \( 
{\mathcal{A}}_{b} \),
while the primary fields of the UV limit of mTh theory are \( 
{\mathcal{A}}_{f} \).

\begin{figure}[h]
{\par\centering \includegraphics{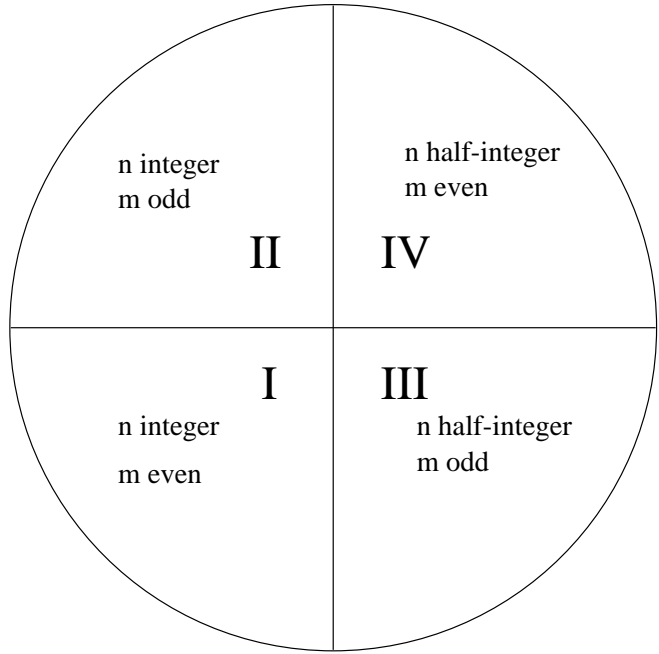} \par}

\caption{{\small 
The family of vertex operators \protect\( V_{(n,m)} \protect\)
with \protect\( n\in \protect\mathbb Z/2 \protect\) and 
\protect\( m\in \protect\mathbb Z\protect\).
Sector \textbf{I} is the common subspace. \textbf{I}
and \textbf{III} are \protect\( 
{\protect\mathcal{A}}_{f} \protect\),
that defines the UV behaviour of massive Thirring; \textbf{I} and
\textbf{II} are \protect\( {\protect\mathcal{A}}_{b}\protect\)
that is the UV of sine-Gordon, \textbf{IV} is a sector of non
mutually local vertex operators.\label{4sectors.eps}} }
\end{figure}

Note that the two algebras share a common subspace with even values of the 
topological
charge, generated by \( \{V_{(n,\, m)}:\, n\in \mathbb {Z},\, m\in 2\mathbb 
{Z}\} \),
where the massive theories described by the Lagrangians (\ref{sG_Lagrangian})
and (\ref{mTh_Lagrangian}) are identical. The well known proof by Coleman about
the equivalence of the two theories \cite{s-coleman} holds exactly in this
subspace. Figure \ref{4sectors.eps} shows the four sectors where all the vertex
operators live.

\section{Light-cone lattice regularization of Sine-Gordon 
theory\label{light-cone.section}}

In this section we present a lattice regularization of the Sine-Gordon model
which is particularly suitable to study FSE. It is well known to lattice 
theorists
that there are many actions on a lattice that all discretize the same continuum
theory. This means that there are many possible regularizations of the same
theory. Lattice researchers are used to choose the action on the lattice 
possessing
the properties that best fit their calculational needs. In the present context
what we would like to keep on a lattice discretization of Sine-Gordon model
is of course the property of integrability. The light-cone lattice construction
we are going to illustrate is a way (not the unique!) to achieve this goal.

\subsection{Kinematics on light-cone lattice}

It is a usual way to regularize quantum field theories by defining them on a
space-like {}``Hamiltonian{}'' lattice (where time is continuous and space
discrete) or space and time-like {}``Euclidean{}'' lattice (when both space
and time are discrete). In statistical mechanics this is not just a 
regularization
method but can be a right microscopic way to describe physical systems. In two
dimensions, the most known approach is to define a rectangular lattice with
axis corresponding to space and time directions and associate to each site an
interaction depending only on the nearest neighboring sites. In this case the
partition function can be expressed in terms of a transfer matrix.

\begin{figure}[h]
{\par\centering \rotatebox{270}{\includegraphics{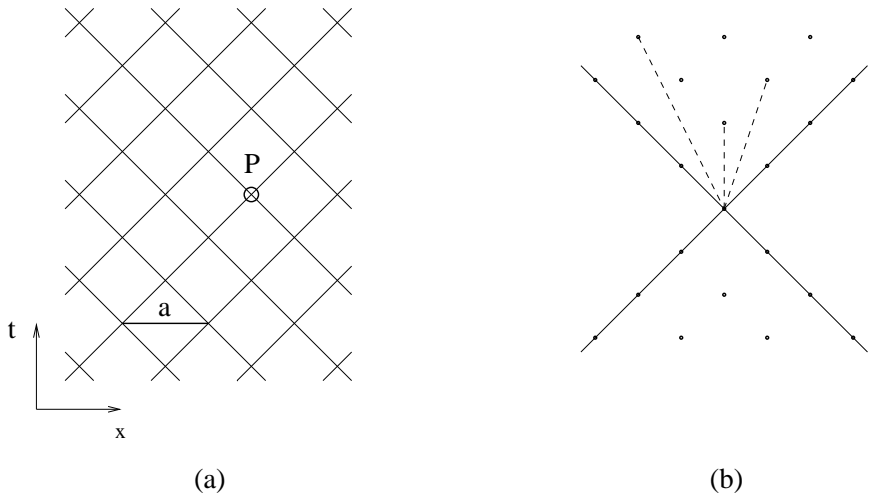}} \par}

\caption{{\small (a) light-cone lattice; (b) double light-cone emanating from 
a point
and high-order interactions (dashed)\label{lcl1.eps}}\small }
\end{figure}

In what follows, a different approach \cite{ddv 87} is adopted: Minkowski and
Euclidean space-time can, in fact, be discretized along light-cone directions.
Light-cone coordinates are: \[
x_{\pm }=x\pm t\]
 and the choice \[
{\mathcal{M}}=\{x_{\pm }=\frac{a}{\sqrt{2}}n_{\pm },\quad n_{\pm }\in \mathbb 
Z\}\]
 defines a light-cone lattice of {}``events{}'' as in figure \ref{lcl1.eps}
(a) . They are spaced by \( a \) in the space and time directions and by \( 
a/\sqrt{2} \)
in light-cone directions. At every event \( P\in {\mathcal{M}} \) a double
light-cone (in the past and in the future) is associated and only events within
this light-cone can be causally connected (see fig. \ref{lcl1.eps} (b)). Then,
any rational and not greater than \( 1 \) speed is permitted for particles,
in an infinite lattice. The shortest displacement of the particle (one lattice
spacing) is realized at speed \( \pm 1 \) and corresponds, from the statistical
point of view, to nearest neighbors interactions. Smaller speeds can be 
obtained
with displacements longer than the fundamental plaquette, and correspond to
high order neighbors interactions. In quantum field theory, these are nonlocal
interactions. 
In the following, only the local case (nearest neighbors) is treated. The 
nearest
neighbors of the event \( P \) are the four nearest points in the light-cone
directions. This implies that particles can have only the speed of light \( 
\pm 1 \)
and are massless. They are called right-movers (R) and left-movers (L).

In this case, it is possible to introduce a useful language for connection with
statistical mechanics associating a particle to a link. Consider the two links
in the future that come out from an event \( P \). Particles R and L in \( P 
\),
by definition, are respectively associated to the right-oriented link and to
the left-oriented link. In this way, the state of a link is defined to be the
state of the point where it begins (also the opposite choice, of connecting
a link with the site where it ends, can be done; it is simply a matter of 
convention).
For example, if in a point \( O \) there is a particle R, one tells that the
{}``right-oriented{}'' link outcoming from it is occupied by R. This 
correspondence
of points and links is possible because only local interactions are assumed,
and it is useful because the counting of states is simpler. In the following,
the lattice is assumed of a finite extent \( L=aN \) in space direction (\( N 
\)
is the number of sites, counted in the space direction), with periodic boundary
conditions, but infinite in time direction. In this way a cylinder topology
is defined for space time. The Hilbert space of states in an event \( P \)
is the tensor product \[
{\mathcal{H}}={\mathcal{H}}_{L}\otimes {\mathcal{H}}_{R}\]
 of R and L space of states. The fact that particles can be classified in left
and right does not mean, in general, that the two dynamics are independent,
as it happens in CFT.

\begin{figure}[h]
{\par\centering \rotatebox{270}{\includegraphics{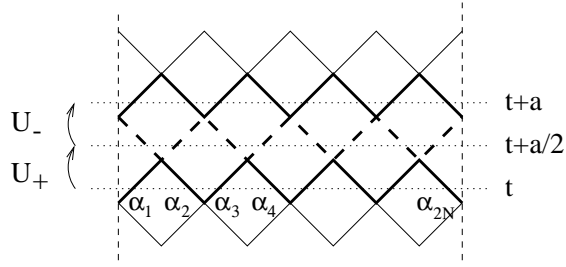}} \par}

\caption{{\small Partial evolution operators. The convention used here for the 
labeling
of states will be used also in the following.\label{upm.eps}}\small }
\end{figure}

Call \( \left| \alpha _{Li},\alpha _{Ri}\right\rangle  \) the generic vector
of a basis of \( {\mathcal{H}}_{i} \) where \( i=1,...,N \) labels the sites.
The notation \[
\left| \alpha _{2i-1},\alpha _{2i}\right\rangle =\left| \alpha _{Li},\alpha 
_{Ri}\right\rangle \]
 is useful and not ambiguous (even number refers to right, odd number refers
to left). The total Hilbert space is: \[
{\mathcal{H}}_{N}=\bigotimes ^{N}_{i=1}{\mathcal{H}}_{i}\]
 and a basic vector can be represented by \[
\left| \alpha _{1},\alpha _{2}\right\rangle \otimes ...\otimes \left| \alpha 
_{2N-1},\alpha _{2N}\right\rangle =\left| \alpha _{1},\alpha _{2},...,\alpha 
_{2N}\right\rangle \in {\mathcal{H}}_{N}.\]
 Note that in a \( N \) sites lattice, due to light-cone, \( 2N \) labels
are required. The state \( \left| \alpha _{1},\alpha _{2},...,\alpha 
_{2N},t\right\rangle  \)
is better thought to lie on links separating two lines of events. The time 
slices
\( t \), \( t+\frac{a}{2} \), \( t+a \) are taken as represented in 
fig.\ref{upm.eps}.
From this figure it is clear that two different types of evolution operators
can be defined, depending on the initial state: \begin{equation}
\label{upm}
\begin{array}{c}
U_{+}\left| \alpha _{1},\alpha _{2},...,\alpha _{2N},t\right\rangle =\left| 
\alpha _{1}',\alpha _{2}',...,\alpha _{2N}',t+a/2\right\rangle \\
\\
U_{-}\left| \alpha _{1}',\alpha _{2}',...,\alpha _{2N}',t+a/2\right\rangle 
=\left| \alpha _{1}'',\alpha _{2}'',...,\alpha _{2N}'',t+a\right\rangle 
\end{array}
\end{equation}
 where the initial states are chosen as in figure \ref{upm.eps}. 
They can be pictorially represented by the links that have to be added to the
time slice \( t \) to get \( t+\frac{a}{2} \) (for \( U_{+} \)) and to \( 
t+\frac{a}{2} \)
to get \( t+a \) (for \( U_{-} \))

\vspace{0.3cm}
{\par\centering \includegraphics{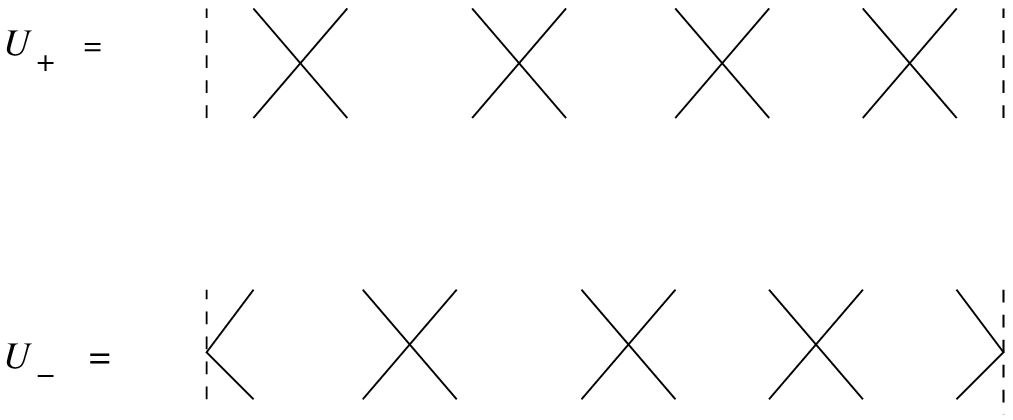} \par}
\vspace{0.3cm}

Schr\"{o}dinger form of equations of motion is used, for a state in Hilbert
space.\footnote{%
In Schrodinger form, if \( \left| \alpha \right\rangle  \) is a state vector,
its time evolution is given by \( \left| \alpha ,t\right\rangle =U\left| 
\alpha ,0\right\rangle  \)
where \( U \) satisfies motion's equations: \( U=Te^{-i\int dtH} \) (Dyson's
series) 
} The global time operator can be chosen as \[
U=U_{+}U_{-\textrm{ }}\: \: \textrm{ or }\: \quad U'=U_{-}U_{+}\]
 depending on the initial state. For a consistent quantum theory, both these
operators must be unitary. This is guaranteed if the assumption \( U^{\dagger 
}_{+}U_{+}=U^{\dagger }_{-}U_{-}=1 \)
is made, that is the elementary operators themselves must be unitary.

\begin{figure}[h]
{\par\centering \rotatebox{270}{\includegraphics{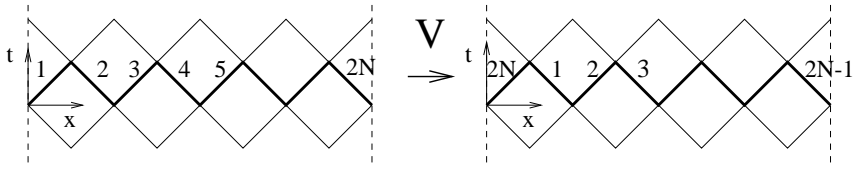}} \par}

\caption{{\small Half-shift operator\label{mezzo-shift.eps}}\small }
\end{figure}

Another operator plays an important role and is defined as follows (the states
are at a certain fixed time): \begin{equation}
\label{mezzoshift}
V\left| \alpha _{1},\alpha _{2},...,\alpha _{2N},t\right\rangle =\left| \alpha 
_{2N},\alpha _{1},...,\alpha _{2N-1},t\right\rangle 
\end{equation}
 It corresponds to an half-space shift in the space direction, with exchange
of right and left states (see the figure \ref{mezzo-shift.eps}).

Two applications of \( V \) give a shift by an entire lattice spacing, then
\( V^{2} \) is the lattice space evolution operator. Also \( V \) is a unitary
operator. By using the pictorial representations introduced above it is easy
to convince oneself that the following relations hold\begin{equation}
\label{comm-uv}
\left[ V^{2},U_{\pm }\right] =0;\qquad U_{\pm }=VU_{\mp }V^{\dagger }.
\end{equation}
For example, start by taking \( U_{-} \). To obtain \( VU_{-} \) one has to
apply \( V \) \emph{after} \( U_{-} \), which means shift the upper endpoints
of \( U_{-} \) by \( \frac{a}{2} \) towards right.

\vspace{0.3cm}
{\par\centering \includegraphics{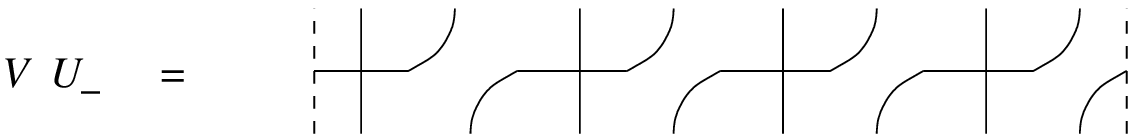} \par}
\vspace{0.3cm}

Now, by applying \( V^{\dagger } \) to the lower part of this drawing, i.e.
shifting the lower endpoints by \( \frac{a}{2} \), one can see that \( 
VU_{-}V^{\dagger }=U_{+} \).
Analogously, all the other relations can be proven.

As a consequence of (\ref{comm-uv}) the two principal unitary evolution 
operators,
\( V^{2} \) and \( U \), are commuting: \[
\left[ V^{2},U\right] =\left[ U,V^{2}\right] =0.\]
 It is natural to identify them as the exponential of the Hamiltonian operator,
and the exponential of the linear momentum: \begin{equation}
\label{unitari}
\begin{array}{c}
U=e^{-iaH}\\
V^{2}=e^{-iaP}.
\end{array}
\end{equation}
 There are other two important operators, defined as: \begin{equation}
\label{unitari-conoluce}
\begin{array}{c}
U_{R}=U_{+}V\\
U_{L}=U_{+}V^{\dagger }
\end{array}
\end{equation}
 They correspond to one step evolution in light-cone directions. They are 
commuting
and give the expressions: \[
\begin{array}{cc}
\quad \quad U=U_{R}U_{L}\quad \quad  & V^{2}=U_{R}U^{\dagger }_{L}\\
\left[ U_{R},U_{L}\right] =0 & U_{R}^{\dagger }U_{R}=U_{L}^{\dagger }U_{L}=1
\end{array}\]
 then, using also (\ref{unitari}) yields: \begin{equation}
\label{unit-lightcone}
U_{R}=e^{-i\frac{a}{2}(H+P)},\qquad U_{L}=e^{-i\frac{a}{2}(H-P)}.
\end{equation}

\subsection{Dynamics on light-cone lattice\label{s-matrix.section}}

A dynamics can be defined by giving all the amplitudes of the different 
processes
that can take place on the lattice. The fundamental assumption is that at every
site a whole process can happen, in the sense that if \( \left| \alpha 
_{L},\alpha _{R}\right\rangle _{in} \)
and \( \left| \beta _{L},\beta _{R}\right\rangle _{out} \) are the incoming
and outgoing states in a certain site, they can be considered asymptotic states
and the transition amplitude is a (bare) S-matrix element. To each site of the
lattice, which is the crossing of 2 incoming and 2 outgoing lines, we associate
a matrix given by the Boltzmann weights of a vertex model \begin{equation}
\label{ampiezza}
\displaystyle _{out}\left\langle \beta _{R},\beta _{L}\right. \left| \alpha 
_{R},\alpha _{L}\right\rangle _{in}=R_{\alpha _{R},\alpha _{L}}^{\beta 
_{R,}\beta _{L}}(\lambda )
\end{equation}
 The parameter \( \lambda  \) is the so called \emph{spectral parameter} of
the \( R \) matrix of vertex models \cite{baxter}. The \( R \) matrix can
be pictorially represented by drawing the crossing where it is put and 
labelling
the external lines with the 4 indices of the matrix

\vspace{0.3cm}
{\par\centering \includegraphics{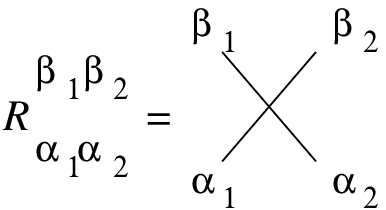} \par}
\vspace{0.3cm}

The system is then defined via its microscopic amplitudes, formally identified
with the Boltzmann weights of some lattice model. If the lattice model is 
integrable,
its Boltzmann weights, in the form of \( R \) matrix, satisfy Yang-Baxter 
equations,
and our bare scattering theory also does. This is the way integrability is 
implemented
into the lattice definition of the model.

From the definition (\ref{upm}) of evolution operators one can see that they
are realized in this context as products of \( R \) matrices\begin{equation}
\label{ups-Smatrix}
\left\langle \alpha _{1},\alpha _{2},...,\alpha _{2N}\right| U_{+}\left| \beta 
_{1},\beta _{2},...,\beta _{2N}\right\rangle =\prod _{i=1}^{N}R_{\alpha 
_{2i-1},\alpha _{2i}}^{\beta _{2i-1},\beta _{2i}}(\lambda -\lambda _{i})
\end{equation}
which, in pictorial form, is nothing else than the already seen definition of
\( U_{+} \), decorated with indices. The operator \( U_{-} \) can be obtained
from (\ref{comm-uv}). The product is on all the sites at a given time. The
particular choice of the spectral parameter, with a {}``correction{}'' 
(inhomogeneity)
depending on the site will become clear in a moment.

Apparently, this {}``phenomenological{}'' approach is quite unusual, because
in traditional lattice quantum field theory every site is associated with an
interaction potential (ex: \( \phi ^{4}(i) \) is the potential on the site
\( i \)), not a whole scattering process. This can appear as a sort of 
{}``macroscopic{}''
approach, not based on fundamental interactions. But the properties of 
factorizable
scattering must be taken into account. Factorization of generic amplitudes in
\( 2\, \rightarrow \, 2 \) particles amplitudes is a sort of quantum 
superposition
principle and the remarkable fact is that between one scattering and the other,
the particles are asymptotic ones, that means that they are free. Every point
contains all the interaction. This is what was assumed in the definition of
the lattice. Then it is perfectly justified that every site is connected with
a whole \( 2\, \rightarrow \, 2 \) particle scattering process.\footnote{%
In the next paragraphs, it will be shown that, for the particular case of the
6 vertex R matrix, a more traditional lattice QFT approach can be formulated,
in terms of a fermionic field. 
}

\subsection{Euclidean transfer matrix}

In order to relate the evolution operators to something computable, we have
to consider the Euclidean lattice vertex model transfer matrix. Consider a two
dimensional Euclidean square lattice, with periodic boundary conditions, in
both the directions. The links are the physical objects of the system. They
can be in different states belonging to the vector spaces \( {\mathcal{A}} \)
(horizontal) and \( {\mathcal{V}} \) (vertical). In principle, such vector
spaces can be different, but in the following they will be assumed as 
isomorphic.
At every site we associate a Boltzmann weight depending on the four links 
crossing
at this site and on a spectral parameter \( \lambda  \) (see fig. \ref{square 
lattice.eps}).
\begin{figure}[h]
{\par\centering \rotatebox{270}{\includegraphics{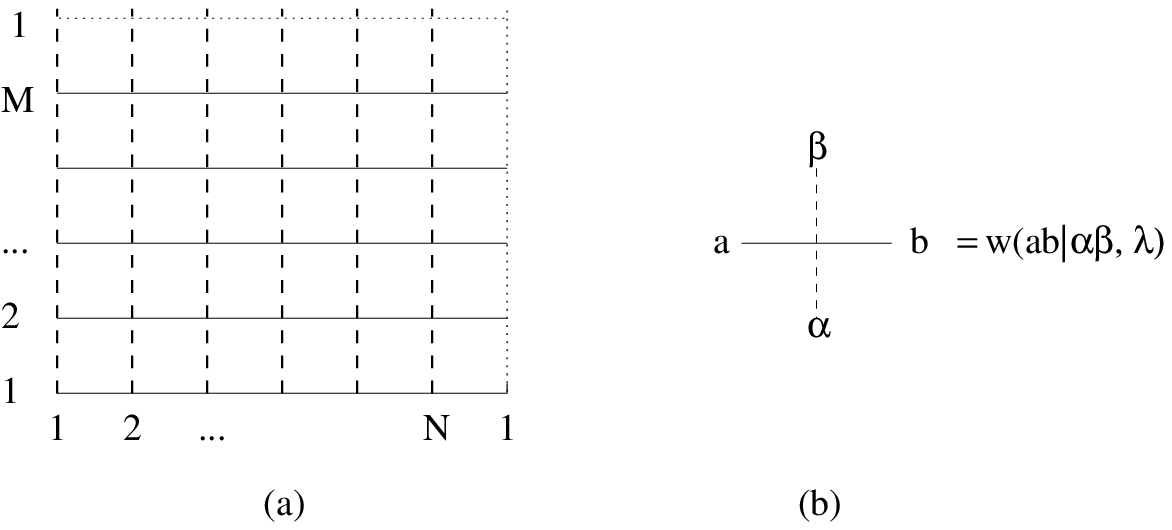}} \par}

\caption{{\small (a) periodic (toroidal) square lattice; (b) Boltzmann weight 
for a
site; the labels are:}\small }

\( a,b\in {\mathcal{A}},\quad \alpha ,\beta \in {\mathcal{V}} \)\label{square 
lattice.eps}
\end{figure}
The simplest case is to take the same Boltzmann weights at all sites, but more
general configurations are possible. For our purposes it will be interesting
to consider the case with an inhomogeneity \( \lambda _{i} \) at each site,
where \( i \) is the column index (i.e. all the sites on a column have the
same inhomogeneity), in the sense that the Boltzmann weight in the column \( i 
\)
is taken to be \( w(ab|\alpha \beta ,\, \lambda -\lambda _{i})=R^{\beta \, 
b}_{a,\alpha }(\lambda -\lambda _{i}) \).
Also for the boundary conditions it is possible to assume more general 
configurations
than the simplest one (i.e. toroidal b.c.). Assume that between the column \( 
N \)
and the \( N \)+1 (that is 1) there is a nontrivial seam line, in such a way
that the Boltzmann weights on the column \( N \) (with respect to the normal
ones) are given by \( e^{i\omega b}R^{\beta \, b}_{a,\alpha }(\lambda -\lambda 
_{N}) \).
This choice is made because only the link \( b \) crosses the seam line. It
implements the so called \emph{twisted boundary conditions} (of course, \( 
\omega =0 \)
reproduces the periodic case).

The object of interest for us is the row-to-row transfer matrix. It is a 
product
of concatenated \( R \) matrices along a whole horizontal line of the lattice
and describes the evolution from one time slice to the next in Euclidean 
lattice.
For simplicity, let us insist on the pictorial notation

\vspace{0.3cm}
{\par\centering \includegraphics{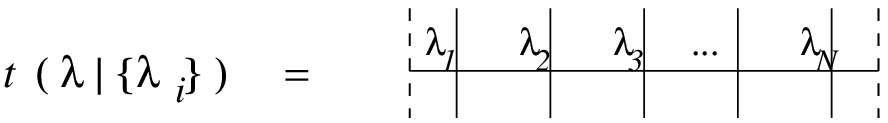} \par}
\vspace{0.3cm}

We have omitted the indices on the two rows to avoid useless heaviness of 
notation.
This object is well known in the statistical mechanics literature and for many
integrable models it can be diagonalized exactly by Bethe Ansatz methods, as
we shall illustrate in next sections. Unfortunately, it refers to a row-to-row
evolution on an Euclidean lattice, and not to the case we are interested in
of an evolution in tine in our light-cone lattice. Is there a way to relate
this known object to the operators \( U_{\pm } \) or \( U_{L,R} \) introduced
above?

The trick comes eventually by considering a particular distribution of 
inhomogeneities.
Compute the transfer matrix at a value \( \Theta  \) of the spectral parameter,
and choose inhomogeneities to be \( \lambda _{i}=(-1)^{i+1}\Theta  \). In this
case, all the \( R \) matrices at odd sites are calculated at spectral 
parameter
\( 2\Theta  \), while those at even sites are calculated at spectral parameter
0. The nice fact is that, for all unitary \( R \) matrices \[
R_{\alpha _{1},\alpha _{2}}^{\beta _{1},\beta _{2}}(0)=\delta _{\alpha 
_{1}}^{\beta _{1}}\delta _{\alpha _{2}}^{\beta _{2}}\]
 or in graphical form

\vspace{0.3cm}
{\par\centering \includegraphics{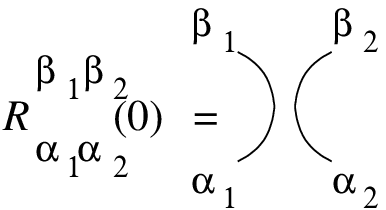} \par}
\vspace{0.3cm}

The transfer matrix \( t(\Theta |\{(-1)^{i+1}\Theta \}) \) then turns out to
be equivalent to the matrix element of the operator \( U_{+} \) taken between
two time slices of the light-cone lattice, as it appears clearly from this 
pictorial
representation 

\vspace{0.3cm}
{\par\centering 
\resizebox*{0.84\textwidth}{0.06\textheight}{\includegraphics{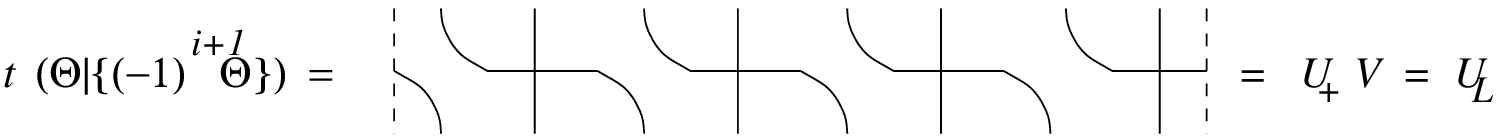}} \par}
\vspace{0.3cm}

Analogously the operator \( U_{R} \) is realized by \( t^{-1}(\Theta 
|\{(-1)^{i+1}\Theta \}) \).
Diagonalization of \( t \) gives therefore the eigenvalues of \( U_{R} \)
and \( U_{L} \) which are the exponentials of \( E+P \) and \( E-P \) 
respectively.
If we are able to compute the eigenvalues of \( t \), we are able to calculate
the energy levels and the total momenta of states in the theory defined on the
light-cone lattice.

\subsection{6 vertex model: main results\label{section:6vertici_varie}}

The theory developed up to now is general and not referred to a specific model.
The simplest non trivial case is the \( 4\times 4 \) R matrix, corresponding
to the choice \( {\mathcal{A}}={\mathcal{V}} \). As shown in \cite{baxter},
the most general solution is the so called 8 vertex model. This name means that
only 8, between the 16 entries of the \( R \) matrix, are nonzero. A special
case is the 6 vertex model, for which many results have been obtained in the
light-cone description: this will be the principal object in the following.
The R matrix has the form (lower index are rows and upper index are columns)
\begin{equation}
\label{6vRmatrix}
R(\vartheta ,\gamma )=\left( \begin{array}{cccc}
a &  &  & \\
 & c & b & \\
 & b & c & \\
 &  &  & a
\end{array}\right) 
\end{equation}

There is a well known mapping between vertex models and spin chains (see 
\cite{baxter}),
i.e. the transfer matrix is the exponential of the quantum Hamiltonian of the
chain: \[
t^{(N)}=e^{-H}.\]
 In the case of 8 vertex model, the Hamiltonian is the XYZ(1/2) chain, while
in the special case of 6 vertex, is the XXZ(1/2) chain. In what follows, this
identification can be useful to interpret some facts connected with Bethe 
Ansatz.
The XXZ(1/2) chain Hamiltonian is given by: \[
H=\sum ^{N}_{i=1}\left[ \sigma _{x}^{i}\sigma _{x}^{i+1}+\sigma _{y}^{i}\sigma 
_{y}^{i+1}+(1-\cos \gamma )\sigma _{z}^{i}\sigma _{z}^{i+1}\right] \]
 and \( \gamma  \) is the anisotropy. The \( \sigma  \) are Pauli matrices.
The total z-component of the spin, which is a conserved quantity in this 
system,
will play an important role in Bethe Ansatz.

It is possible to give a particle interpretation to the same \( R \) matrix
(\ref{6vRmatrix}) on the light-cone lattice. The simplest case of particles
obeying Pauli exclusion principle (fermions) and without internal degrees of
freedom (color number) is assumed. This means that in an event only one 
particle
of type R and one L at most can take place. In other words, one link has two
states: empty or occupied. At every point there are four links, that means 16
possible configurations associated to it. In terms of events, these are the
possible configurations that connect a point with the nearest neighbors in the
future.

\begin{figure}[h]
{\par\centering 
\resizebox*{0.84\textwidth}{0.1\textheight}{\rotatebox{270}{\includegraphics{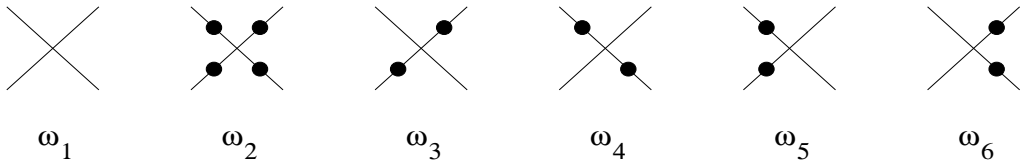}}} 
\par}

\caption{{\small 6 permitted amplitudes. Dots are 
particles.\label{amplitude.eps}}\small }
\end{figure}

Assume now that only amplitudes that conserve the total number of particles
(R+L) are non vanishing. This reduces to 6 the permitted configurations, as
it is shown in the figure \ref{amplitude.eps}. This is simply the 6 vertex
model whose \( R \) matrix is written in (\ref{6vRmatrix}).
The assumption of integrability for this amplitudes gives the general 
six-vertex
model. The requirement of symmetry under parity transformation implies that
\( \omega _{3}=\omega _{4} \) and \( \omega _{5}=\omega _{6} \). The convention
adopted in (\ref{6vRmatrix}) gives \[
b=\omega _{3}=\omega _{4},\quad c=\omega _{5}=\omega _{6}.\]
 This R matrix can be written in an operatorial form, by defining a lattice
chiral fermion \( \psi _{A,n} \), with \( A=R,L \), and \( n \) labels the
sites. The anticommutation rules are the canonical ones: \begin{equation}
\label{fermion}
\left\{ \psi _{A,n},\psi _{B,m}\right\} =0,\qquad \left\{ \psi _{A,n},\psi 
^{\dagger }_{B,m}\right\} =\delta _{AB}\delta _{nm}.
\end{equation}
 This fermion has some interesting properties, that are exposed in the paper
\cite{ddv 87}, and are sketched in the following list:

\begin{enumerate}
\item the R matrix and all the other operators \( U_{any} \) can be written in 
an
operatorial form in terms of the fermion; 
\item the lattice Hamiltonian, in the free case \( \omega _{1}=\omega 
_{2}=b=1,\; c=0 \),
can be explicitly written; by this, the following dispersion relation can be
obtained: \( E=\pm k \); this is the dispersion relation for a free massless
particle; the unusual fact is that it is monotonous, then there is no doubling
of fermions; this is a consequence of the non-locality of the Hamiltonian 
\item the lattice Hamiltonian admits a continuum limit \( N\, \rightarrow \, 
\infty  \)
and \( a\, \rightarrow \, 0 \) but with \( L=Na \) fixed; locality is recovered
in this limit; the continuum equations of motion are those of the massive 
Thirring
model. The space is compactified on a cylinder. 
\end{enumerate}
In the continuum limit, the massive Thirring model emerges as the field theory
characterizing the scaling behavior of the dynamics on the lattice (remember
that \( L \) is finite; the scaling behaviour is understood in terms of this
\( L \)). We conclude that the theory we have constructed is a lattice 
regularization
of the Thirring model suitable to study its integrability properties.

\subsection{6 vertex model: Bethe Ansatz\label{section:6_vertex_BA}}

The assumptions of unitarity and hermitian analyticity will be taken into 
account,
for the \( R \) matrix, and this requires that the variables in 
(\ref{6vRmatrix})
must have the specific form: \begin{equation}
\label{R_{m}atrix_{e}ntries}
a=a(\vartheta ,\gamma )=\sinh (i\gamma -\vartheta ),\quad b=b(\vartheta 
,\gamma )=\sinh \vartheta ,\quad c=c(\vartheta ,\gamma )=i\sin \gamma 
\end{equation}

The transfer matrix defined by this \( R \) matrix can be diagonalized with
Bethe Ansatz method \cite{bethe}. In terms of the spin chain, this means that
there are two operators, usually indicated by \( B(\vartheta ) \) and \( 
C(\vartheta ) \),
whose expression is known, and there is a {}``reference state{}''\footnote{%
The {}``reference state{}'' here is only a mathematical object. Physically
speaking, it corresponds to the ferromagnetic state with all the spins up. 
} \( \left| \Omega \right\rangle  \) such that: \[
C(\vartheta )\left| \Omega \right\rangle =0\]
 and \begin{equation}
\label{base}
B(\vartheta _{1})...B(\vartheta _{M})\left| \Omega \right\rangle ,
\end{equation}
 for appropriate values of \( \vartheta _{j} \), is an eigenstate of the 
transfer
matrix. The {}``appropriate values{}'' of \( \vartheta _{j} \) can be obtained
as the solution of a set of \( M \) coupled nonlinear equations, that are 
called
Bethe Ansatz equations. In general, the transfer matrix contains all the 
conserved
charges, in particular the Hamiltonian. Then the Bethe Ansatz eigenstates are
also eigenstates of the Hamiltonian. The Hilbert space of the theory and the
action of conserved charges on it are then exactly known.

All this computations for the 6 vertex model were performed in \cite{baxter, 
fadd-takt, devega 89};
the final results are written here for the eigenvalues \( \tau (\vartheta 
,\Theta ,\omega ) \)
of the inhomogeneous and twisted transfer matrix (here \( \vartheta  \) is
the spectral parameter, \( \Theta  \) the inhomogeneity and \( \omega  \)
the twist) \[
\begin{array}{c}
\displaystyle \tau (\vartheta ,\Theta ,\omega )=e^{i\omega }\left[ a(\vartheta 
-\Theta )\, a(\vartheta +\Theta )\right] ^{N}\prod ^{M}_{j=1}\frac{\sinh 
\frac{\gamma }{\pi }\left[ i\frac{\pi }{2}+\vartheta _{j}+\vartheta \right] 
}{\sinh \frac{\gamma }{\pi }\left[ i\frac{\pi }{2}-\vartheta _{j}-\vartheta 
\right] }+\\
\displaystyle +e^{-i\omega }\left[ b(\vartheta -\Theta )\, b(\vartheta +\Theta 
)\right] 
^{N}\prod ^{M}_{j=1}\frac{\sinh \frac{\gamma }{\pi }\left[ i\frac{3\pi 
}{2}-\vartheta _{j}-\vartheta \right] }{\sinh \frac{\gamma }{\pi }\left[ 
-i\frac{\pi }{2}+\vartheta _{j}+\vartheta \right] }
\end{array}\]
 and the values of \( \vartheta _{j} \) are defined by the set of coupled 
nonlinear
equations called Bethe Ansatz equations: \[
\begin{array}{c}
\label{bethe}
\displaystyle \left( \frac{\sinh \frac{\gamma }{\pi }\left[ \vartheta 
_{j}+\Theta +\frac{i\pi }{2}\right] \sinh \frac{\gamma }{\pi }\left[ \vartheta 
_{j}-\Theta +\frac{i\pi }{2}\right] }{\sinh \frac{\gamma }{\pi }\left[ 
\vartheta _{j}+\Theta -\frac{i\pi }{2}\right] \sinh \frac{\gamma }{\pi }\left[ 
\vartheta _{j}-\Theta -\frac{i\pi }{2}\right] }\right) ^{N}=\\
=\displaystyle -e^{2i\omega 
}\prod _{k=1}^{M}\frac{\sinh \frac{\gamma }{\pi }\left[ \vartheta 
_{j}-\vartheta _{k}+i\pi \right] }{\sinh \frac{\gamma }{\pi }\left[ \vartheta 
_{j}-\vartheta _{k}-i\pi \right] }
\end{array}\]
 where \( 2N \) is the length of the chain and \( N \) the number of sites
in a row of the light-cone lattice. The \( \vartheta _{j} \) are called 
\emph{Bethe
roots}, and in principle, can take any complex value. But there is a 
periodicity
in their imaginary part: \begin{equation}
\label{periodicita'}
\displaystyle \vartheta _{j}\, \rightarrow \, \vartheta _{j}+\frac{\pi 
^{2}}{\gamma }i
\end{equation}
 then only a strip around the real axis must be taken into account for the 
Bethe
roots: \begin{equation}
\label{strisciafisica}
\vartheta _{j}\in \mathbb {R}\times i\, \left] -\frac{\pi ^{2}}{2\gamma 
},\frac{\pi ^{2}}{2\gamma }\right] .
\end{equation}
 Moreover, only the range \( 0<\gamma <\pi  \) will be examined.

The whole spectrum of the theory can be obtained using all the Bethe 
configurations
having \( M\leq N \) and \( \vartheta _{j}\neq \vartheta _{k} \) for every
\( j\neq k. \) In general for a state with \( M \) roots the third component
of the spin of the chain is \begin{equation}
\label{spin-chain}
S=N-M,
\end{equation}
 because every operator \( B(\vartheta _{i}) \) counts as \( -1 \) spin.

This XXZ(1/2) chain has 2 states in every site, then \( 2^{2N} \) states in
total. Then the energy spectrum is upper and lower bounded. Changing the sign
of the Hamiltonian gives another permitted physical system. This means that
there are two possible vacua. The first one is the so called ferromagnetic 
ground
state, corresponding to \( M=0 \) that is the reference state \( \left| \Omega 
\right\rangle . \)
It has spin \( S=N \). The second one is the antiferromagnetic ground state,
that can be obtained with \( M=N \) and all the roots \( \vartheta _{i} \)
real. It has spin \( S=0 \).

In what follows, only the antiferromagnetic ground state will be considered,
because it has one important property: in the thermodynamic limit (\( N\, 
\rightarrow \, \infty  \))
it can be interpreted as a Dirac vacuum (a sea of particles created by \( B \))
and the excitations on this vacuum behave as particles.

The energy \( E \) and momentum \( P \) of a state can be read out by the
transfer matrix eigenvalues. The final form is:

\begin{equation}
\label{autovalori}
\displaystyle e^{i\frac{a}{2}(E\pm P)}=e^{\pm i\omega }\prod 
^{M}_{j=1}\frac{\sinh \frac{\gamma }{\pi }\left[ i\frac{\pi }{2}-\Theta \pm 
\vartheta _{j}\right] }{\sinh \frac{\gamma }{\pi }\left[ i\frac{\pi 
}{2}+\Theta \mp \vartheta _{j}\right] }.
\end{equation}
 Other integrals of motion can be obtained in a similar way by transfer matrix.
Note that the second term in the transfer matrix expression vanishes because
\( b(0)=0 \).

\section{Nonlinear Integral Equation from Bethe Ansatz}

In this section the fundamental nonlinear integral equation driving sG scaling
functions is derived. In the literature it is known as Destri-de Vega equation.
It was obtained first in \cite{ddv 95,klumper} for the vacuum scaling function.
The treatment of excited states was pioneered in \cite{fioravanti} and refined
in \cite{ddv 97,noi PL1}, to arrive to the final form, in \cite{noi 
NP,tesi-Giovanni}.

\subsection{Counting function\label{section:count-funct}}

It is possible to write the Bethe equations (\ref{bethe}) in terms of a 
\emph{counting
function} \( Z_{N}(\vartheta ). \) 

First, introduce the function\footnote{%
In the following we use the parameter \( p=\frac{\pi }{\gamma }-1 \). It will
later turn out to be exactly the same as the one introduced in 
(\ref{parametro_p}).
} \[
\phi _{\nu }(\vartheta )=i\log \frac{\sinh \frac{1}{p+1}(i\pi \nu +\vartheta 
)}{\sinh \frac{1}{p+1}(i\pi \nu -\vartheta )},\qquad \phi _{\nu }(-\vartheta 
)=-\phi _{\nu }(\vartheta )\]
 The oddness on the analyticity strip around the real axis defines a precise
choice of the logarithmic branch. The counting function is defined by 
\begin{equation}
\label{def.Zn}
\displaystyle Z_{N}(\vartheta )=N[\phi _{1/2}(\vartheta +\Theta )+\phi 
_{1/2}(\vartheta -\Theta )]-\sum _{k=1}^{M}\phi _{1}(\vartheta -\vartheta 
_{k})+2\omega 
\end{equation}
 The logarithm of the Bethe equations boils down to the simple condition 
\begin{equation}
\label{quantum}
\displaystyle Z_{N}(\vartheta _{j})=2\pi I_{j}\, ,\quad I_{j}\in \mathbb 
{Z}+\frac{1+\delta }{2},\quad \delta =(M)_{mod\, 2}=(N-S)_{mod\, 2}\in 
\left\{ 0,1\right\}
\end{equation}
 The number \( I_{j} \) plays the role of a quantum number for the Bethe basic
vectors (\ref{base}) and it must be chosen depending on the value of \( \delta 
. \)
Notice that \( \delta  \) and \( \omega  \) play a similar role, because both
produce a shift in the quantum numbers \( I_{j} \) (if \( \omega  \) is 
absorbed
in the definition of \( I_{j} \)): in the first case the shift is exactly \( 
\pi , \)
in the second case it is a real (possibly complex) number. This means that the
variable \( \delta  \) can be absorbed in \( \omega  \) but the most convenient
choice is to use them both.

Observe that Bethe roots can be obtained as zeros of the equation: 
\begin{equation}
\label{zero-condition}
1+(-1)^{\delta }e^{iZ_{N}(\vartheta _{j})}=0
\end{equation}

\subsection{Classification of Bethe roots and counting 
equation\label{classificazione}}

From Bethe Ansatz it is known that a Bethe state (\ref{base}) is uniquely 
characterized
by the set of quantum numbers \( \left\{ I_{j}\right\} _{j=1,...,M}\: ,\quad 
M\leq N \)
that appear in (\ref{quantum}). Notice that \( M\leq N \) means \( S\geq 0. \)
The values of \( \vartheta _{j} \) to put in (\ref{base}) can be obtained
solving Bethe equations. It is also known that only states with \[
\vartheta _{j}\neq \vartheta _{i},\quad \forall \: j\neq i\]
 are allowed. It is a sort of fermionic character for Bethe states 
\cite{faddeev 95}.

Bethe roots can either be real or appear in complex conjugate pairs. In the
specific case (\ref{bethe}), there is another possibility, due to periodicity
(\ref{periodicita'}): if a complex solution has imaginary part \( Im\, 
\vartheta =\frac{\pi }{2}(p+1) \)
it appears as a single root (in (\ref{bethe}) it produces the left hand side
real, then its complex conjugate is not required). It is called 
\emph{self-conjugate
root}. Remember now that the maximal number of real roots (\( M=N \)) describes
the antiferromagnetic ground state. From the point of view of the counting 
function,
a more precise classification of roots is required:

\begin{itemize}
\item \emph{real roots}; they are real solutions of (\ref{quantum}) that 
appear in
the vector (\ref{base}); their number is \( M_{R} \); 
\item \emph{holes}; real solutions of (\ref{quantum}) that do NOT appear in 
the vector
(\ref{base}); their number is \( N_{H} \); 
\item \emph{special roots/holes} (special objects); they are real roots or 
holes where
the derivative \( Z_{N}'(\vartheta _{j}) \) is negative\footnote{%
The characteristics of this type of solutions will be clarified in the next
sections.
}; their number is \( N_{S} \); they must be counted both as {}``normal{}''
and as {}``special{}'' objects; 
\item \emph{close pairs}; complex conjugate solutions with imaginary part in 
the range
\( 0<|I m\, \vartheta |<\pi \min (1,p) \); their number is \( M_{C} \); 
\item \emph{wide roots in pairs}; complex conjugate solutions with imaginary 
part
in the range \( \pi \min (1,p)<|I m\, \vartheta |<\pi \frac{p+1}{2} \); 
\item \emph{self-conjugate roots} (wide roots appearing as single); \( I m\, 
\vartheta =\pi \frac{p+1}{2} \);
their number is \( M_{SC} \).
\end{itemize}
The total number of wide roots appearing in pairs or as single is \( M_{W} \).
The following notation will be used (sometimes) for later convenience, to 
indicate
the position of the solutions: \( h_{j} \) for holes, \( y_{j} \) for special
objects, \( c_{j} \) for close roots, \( w_{j} \) for wide roots.

\begin{figure}[h]
{\par\centering 
\resizebox*{0.9\textwidth}{0.37\textheight}{\includegraphics{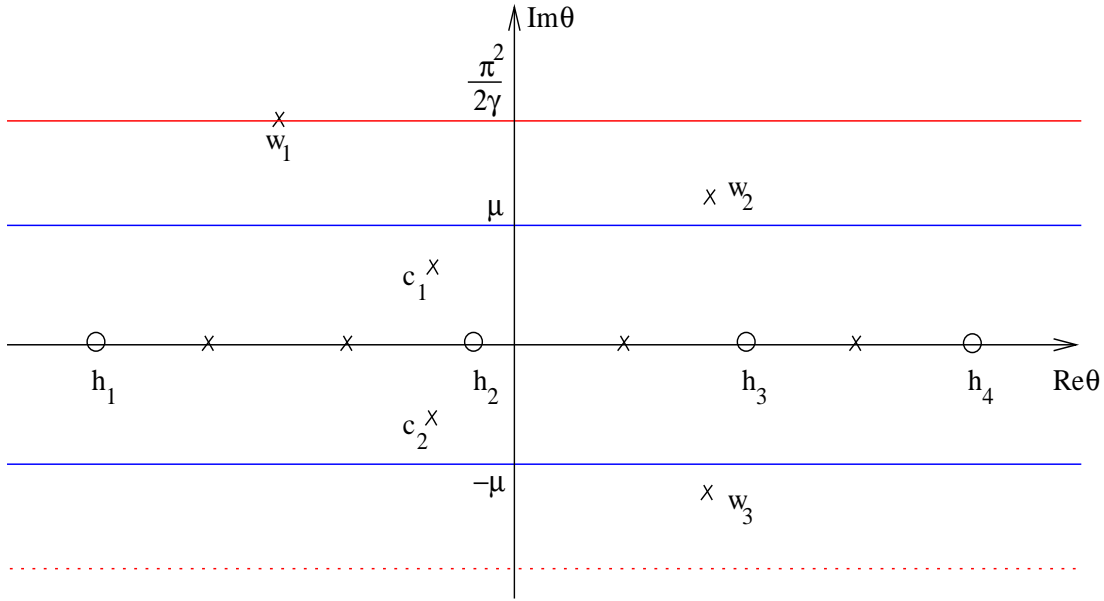}} 
\par}

\caption{{\small
The different types of roots and holes and their position in the complex 
plane. \protect\( \mu \protect\) denotes \protect\( 
\pi \min (1,p) \protect\).
The upper line at \protect\( \protect\frac{\pi ^{2}}{2\gamma }=
\protect\frac{\pi}{2}(p+1)\protect\)
is the self-conjugate one.\label{radici.eps}}\small }
\end{figure}

Complex roots with imaginary part larger than the self-conjugates are not 
required
because of the periodicity of Bethe equations. A graphical representation of
the various types of solutions is given in figure \ref{radici.eps}. 

An important remark must be done: from the definition of \( Z_{N} \) 
(\ref{def.Zn})
it is obvious that only for states without complex roots the fundamental strip
for \( \phi _{\nu }(\vartheta ) \)), that is the largest strip around the real
axis without singularities, is the fundamental strip for \( Z_{N}. \) In all
the other cases the analyticity strip for \( Z_{N} \) is narrower, and depends
on the imaginary parts of the complex roots.

An important property follows from this classification: the \( Z_{N} \) 
function
is \emph{real analytic} if \( \omega  \) is a real number \begin{equation}
\label{analiticita_{r}eale}
Z_{N}\left( \vartheta ^{*}\right) =\left( Z_{N}(\vartheta )\right) ^{*}
\end{equation}

By considering asymptotic values of \( \phi _{\nu }(\vartheta ) \) and \( 
Z_{N}(\vartheta ) \)
for \( \vartheta \rightarrow \pm \infty  \), it is possible to obtain an 
equation
relating the numbers of all the various types of roots. We refer the reader
interested in the details of the derivation to \cite{ddv 97,tesi-Giovanni}.
Here we only mention the final result, in the form where the continuum limit
\( N\rightarrow \infty  \), \( a\rightarrow 0 \) and \( L=Na \) finite, is
already taken\begin{equation}
\label{counting-eq}
N_{H}-2N_{S}=2S+M_{C}+2\, \theta (p-1)\, M_{W}
\end{equation}
 where \( \theta (x) \) is the step function: \( \theta (x)=0 \) for \( x<0 \)
and \( \theta (x)=1 \) for \( x>0 \). Recall that \( S \) is a nonnegative
integer. In the case of \( \omega =0 \), it turns out that \( N_{H} \) is
even (\( M_{C} \) is the number of close roots, and is even).

The most important fact is that the number of real roots does not appear in
this equation . This fact, together to what will be explained in the next 
paragraph,
allows to consider the real roots as a sea of particles (Dirac vacuum) and all
other types of solutions (holes, complex) as excitations on this sea.

\subsection{Non linear integral equation\label{section:NLIE_1}}

\begin{figure}[h]
{\par\centering \includegraphics{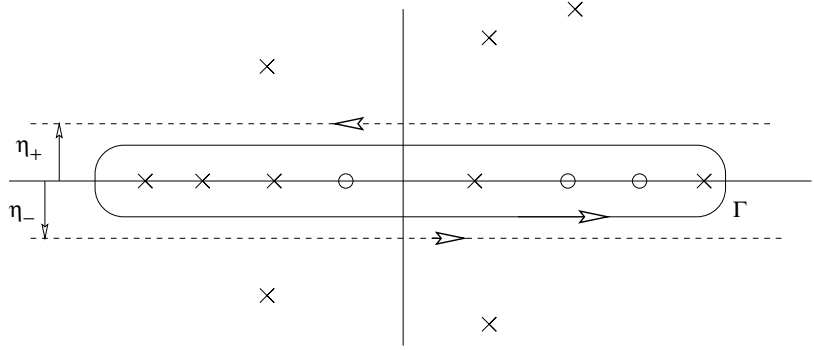} \par}

\caption{{\small Contours for the integration. The crosses are roots while the 
circles
are holes. \label{curvagamma.eps}}\small }
\end{figure}

\noindent Let \( \hat{x} \) be a real solution of the Bethe equation. Thanks
to Cauchy theorem, an analytic function \( f(x) \) on an appropriate strip
containing the real axis admits the following representation \begin{equation}
\label{cauchy}
\displaystyle f(\hat{x})=\oint _{\Gamma _{\hat{x}}}\frac{d\mu }{2\pi 
i}\frac{f(\mu )}{\mu -\hat{x}}=\oint _{\Gamma _{\hat{x}}}\frac{d\mu }{2\pi 
i}f(\mu )\frac{(-1)^{\delta }e^{iZ_{N}(\mu )}iZ_{N}'(\mu )}{1+(-1)^{\delta 
}e^{iZ_{N}(\mu )}}
\end{equation}
 where \( \Gamma _{\hat{x}} \) is a anti-clockwise curve encircling \( \hat{x} 
\)
and avoiding other singularities of the denominator, i.e. other Bethe solutions
(real or complex). In the region where \( \phi _{1}(\vartheta ) \) is analytic,
we can use (\ref{cauchy}) to write

\begin{equation}
\label{integr_{g}amma}
\begin{array}{c}
\displaystyle \sum ^{M_{R}+N_{H}}_{k=1}\phi _{1}(\vartheta -x_{k})=\sum 
^{M_{R}+N_{H}}_{k=1}\oint _{\Gamma _{x_{k}}}\frac{d\mu }{2\pi i}\phi 
_{1}(\vartheta -\mu )\frac{(-1)^{\delta }e^{iZ_{N}(\mu )}iZ_{N}'(\mu 
)}{1+(-1)^{\delta }e^{iZ_{N}(\mu )}}=\\
\displaystyle =\oint _{\Gamma }\frac{d\mu }{2\pi i}\phi _{1}(\vartheta -\mu 
)\frac{(-1)^{\delta }e^{iZ_{N}(\mu )}iZ_{N}'(\mu )}{1+(-1)^{\delta 
}e^{iZ_{N}(\mu )}}
\end{array}
\end{equation}
 The sum on the contours was modified to a unique curve \( \Gamma  \) 
encircling
all the real solutions \( {x_{k}} \), and avoiding the complex Bethe solutions
(this is possible because they are finite in number), as in the figure 
\ref{curvagamma.eps}.

Clearly the \( \Gamma  \) curve must be contained in the strip \[
0<\eta _{+},\eta _{-}<\min \{\pi ,\pi p,|I m\: c_{k}|\: \forall \: k\}\]
 Without loss of generality, assume that \( \eta _{+}=\eta _{-}=\eta  \), and
deform \( \Gamma  \) to the contour of the strip characterized by \( \eta  \).
The regions at \( \pm \infty  \) do no contribute because of the vanishing
of \( \phi ' \), then the integral can be performed on the lines \( \mu =x\pm 
i\eta  \),
where \( x \) is real. After algebraic manipulations involving integrations
by parts and convolutions (for details see \cite{tesi-Giovanni}) one arrives
at a non-linear integral equation (NLIE) to be satisfied by \( Z_{N}(\vartheta 
) \)
\textcolor{black}{\[
\begin{array}{rl}
\displaystyle Z_{N}(\vartheta ) & 
\displaystyle =2N\arctan \frac{\sinh \vartheta }{\cosh 
\Theta }+\sum ^{N_{H}}_{k=1}\chi (\vartheta -h_{k})-2\sum ^{N_{S}}_{k=1}\chi 
(\vartheta -y_{k})-\\
 & \displaystyle -\sum ^{M_{C}}_{k=1}\chi (\vartheta -c_{k})
 -\sum ^{M_{W}}_{k=1}\chi (\vartheta -w_{k})_{II}+\\
 & \displaystyle +2 Im \int ^{\infty 
}_{-\infty }d\rho G(\vartheta -\rho -i\eta )\log \left( 1+(-1)^{\delta 
}e^{iZ_{N}(\rho +i\eta )}\right) 
\end{array}\]
The kernel \begin{equation}
\label{funzioneG}
G(\vartheta )=\frac{1}{2\pi }\int ^{+\infty }_{-\infty }dk\, e^{ik\vartheta 
}\frac{\sinh \frac{\pi (p-1)k}{2}}{2\sinh \frac{\pi pk}{2}\, \cosh \frac{\pi 
k}{2}}\, ,\, p=\frac{\pi }{\gamma }-1
\end{equation}
presents a singularity at the same place where \( \phi _{1}(\vartheta ) \)
does. Any analytic continuation outside the fundamental strip} \( 0<|I m\, 
\vartheta |<\pi \min (1,p) \)
\textcolor{black}{(I determination region) must take this fact into account.
The source terms are given by \[
\chi (\vartheta )=2\pi \int _{0}^{\vartheta }dxG(x)\]
and \[
\chi (\vartheta )_{II}=\left\{ \begin{array}{ll}
\chi (\vartheta )+\chi \left( \vartheta -i\pi \mathrm{sign}\left( I 
m\vartheta \right) \right) \: , & p>1\: ,\\
\chi (\vartheta )-\chi \left( \vartheta -ip\pi \mathrm{sign}\left( I 
m\vartheta \right) \right) \: , & p<1\: .
\end{array}\right. \]
is a modification of the source term due to the analytic continuation over the
strip} \( 0<|I m\, \vartheta |<\pi \min (1,p) \), i.e. in the so called II
determination region.

Such equation is equivalent to the Bethe Ansatz, in the sense that, once 
solved,
it gives the counting function from which the Bethe roots can be reconstructed.
Once the Bethe roots are known, one can use them into eqs.(\ref{autovalori})
to compute the energy and momentum for a given state.

\subsection{Continuum limit\label{section:limite_cont.}}

Although such NLIE is already a precious tool for the lattice model itself,
its importance becomes essential when a continuum limit is done.

The continuum limit must be performed carefully in the light-cone lattice 
setup.
As already mentioned, one has to take \( N\rightarrow \infty  \) (the so called
thermodynamic limit of Statistical mechanics) and the \( a\rightarrow 0 \)
limit simultaneously, in such a way that the product \( L=Na \) keeps finite.
This is the way to implement finite size in the continuum theory, removing all
lattice artifacts but without destroying the cylindrical geometry. However,
one can convince himself, by performing calculations explicitly, that if the
limit is taken by keeping the \( \Theta  \) parameter fixed, the lattice NLIE
blows up to infinity and looses meaning. This reflects the fact that the number
of roots, and hence of Bethe equations goes as \( N \) and becomes infinite
in the thermodynamic limit, thus becoming intractable. However, as already 
forecast
in ref.\cite{ddv 89}, if one assumes a dependence of \( \Theta  \) on \( N \)
of the form\begin{equation}
\label{teta_{n}}
\displaystyle \Theta \approx \log \frac{4N}{{\mathcal{M}}L}.
\end{equation}
it is possible to get a finite limit of the lattice NLIE. This limit is exactly
the one that brings the lattice fermion introduced in 
\ref{section:6vertici_varie}
into the Thirring fermion on the continuum. It is therefore very natural to
give meaning to the continuum NLIE thus obtained as the remnant of Bethe 
Equations
for the renormalized QFT. Notice that sending \( \Theta \rightarrow \infty  \)
in this way naturally introduces a renormalized physical mass \( {\mathcal{M}} 
\).

The \emph{continuum counting function} is defined by: \begin{equation}
\label{def_{Z}}
\displaystyle Z(\vartheta )=\lim _{N\, \rightarrow \, \infty }Z_{N}(\vartheta 
).
\end{equation}
 We present here the continuum NLIE in its full generality

\textcolor{black}{\begin{equation}
\label{nlie-cont}
\displaystyle Z(\vartheta )=l\sinh \vartheta +g(\vartheta |\vartheta 
_{j})+\alpha +2 Im \int ^{\infty }_{-\infty }dxG(\vartheta -x-i\eta )\log 
\left( 1+(-1)^{\delta }e^{iZ(x+i\eta )}\right) 
\end{equation}
where \( l={\cal M}L \). The source term} 
\textcolor{black}{\emph{}}\textcolor{black}{\( g(\vartheta |\vartheta _{j}) \)
is \[
\displaystyle g(\vartheta |\vartheta _{j})=\sum ^{N_{H}}_{k=1}\chi (\vartheta 
-h_{k})-2\sum ^{N_{S}}_{k=1}\chi (\vartheta -y_{k})-\sum ^{M_{C}}_{k=1}\chi 
(\vartheta -c_{k})-\sum ^{M_{W}}_{k=1}\chi (\vartheta -w_{k})_{II}\]
The positions of \( \{\vartheta _{j}\}\equiv \{h_{j}\, ,\, y_{j}\, ,\, c_{j}\, 
,\, w_{j}\} \)
are fixed by the Bethe quantization conditions\[
Z(\vartheta _{j})=2\pi I_{j}\quad ,\quad I_{j}=\mathbb Z+\frac{1+\delta }{2}\]
The parameter \( \delta  \) can be both 0 or 1 in principle. On the lattice
it was determined by the total number of roots, which has become infinite here
together with \( N \). We shall see that locality implements restrictions on
\( \delta  \). The vacuum corresponds to a choice \( \delta =0 \). The 
parameter
\( \alpha  \) takes into account the twist}\begin{equation}
\label{alfa}
\displaystyle \alpha =\omega \, \frac{p+1}{p}+\pi\frac{p-1}{2p}\left( 
\left\lfloor \frac{1}{2}+\frac{S}{p+1}+\frac{\omega }{\pi }\right\rfloor 
-\left\lfloor \frac{1}{2}+\frac{S}{p+1}-\frac{\omega }{\pi }\right\rfloor 
\right) 
\end{equation}
In (\ref{bethe}) the twist term \( \omega  \) is invariant for the shift \[
\omega \, \rightarrow \, \omega +\pi .\]
 The same sort of invariance is required in \( \alpha  \) for the NLIE (which
is equivalent to Bethe equations). It is simple to verify that the expression
for \( \alpha  \) (\ref{alfa}) displays the following symmetry \[
\alpha \, \rightarrow \, \alpha +2\pi \, \, \, \textrm{when}\, \, \, \omega \, 
\rightarrow \, \omega +\pi \, .\]
 Note that shifting \( \alpha  \) by \( 2\pi  \) is an invariance of the NLIE
(\ref{nlie-cont}), if an appropriate redefinition of the Bethe quantum numbers
is made: \( I_{j}\, \rightarrow \, I_{j}+1 \). This shift does not affect 
physical
quantities, that depend only on the variables \( \vartheta _{j} \). 
\textcolor{black}{The
untwisted case corresponds to \( \alpha =0 \).}

\textcolor{black}{In practical calculations, one starts from a \( Z(\vartheta 
) \)
without any roots other than the real ones (i.e. the counting function of the
vacuum). This will be a solution of (\ref{nlie-cont}) with \( g(\vartheta 
|\vartheta _{j})=0 \).
It can be got by numerical iteration starting from the driving term \( l\sinh 
\vartheta  \).
The numerical convergence is normally quite fast. It is more efficient for 
large
\( l \), while it becomes a bit slower in the region near \( l\rightarrow 0 \).
Once the vacuum \( Z(\vartheta ) \) is determined, one can start fixing the
number of roots, holes, etc... that must appear in a particular state he wants
to study. This configuration must, of course, be compatible with the counting
equation (\ref{counting-eq}). The first trial positions for \( \vartheta 
_{j}\in \{h_{j}\, ,\, y_{j}\, ,\, c_{j}\, ,\, w_{j}\} \)
will be determined by the quantization conditions \( Z(\vartheta _{j})=2\pi 
I_{j} \).
The (integers or half-integers) numbers \( I_{j} \) are chosen in principle
freely, but it turns out that some of them are constrained in terms of the 
others.
In such a way a first trial of \( g(\vartheta |\vartheta _{j}) \) is created
and put into the NLIE. Iteration is redone, in order to get a better 
approximation
for \( Z(\vartheta ) \), from which a better determination of the positions
of \( \vartheta _{j} \) can be obtained again by using the quantization rules,
which in turn gives a better \( g(\vartheta |\vartheta _{j}) \), and so on
so forth, up to convergence. All this procedure can be done in few minutes of
computing on a typical Linux/Intel platform without resorting to any 
supercomputer
or other technically advanced tool.}

\subsection{\textcolor{black}{Energy and momentum expressions}}

With a procedure analogous to the one sketched above, i.e. apply the residue
trick and then do a continuum limit along the lines of the previous section,
it is possible to give integral expressions for the energy and momentum. 
Starting
from (\ref{autovalori}) one can isolate an extensive (i.e. proportional to
\( N \)) bulk term to be subtracted on the continuum. The remaining finite
part of the energy can be written in a form where the positions of the various
kinds of roots and the function \( Z(\vartheta ) \) are the only 
inputs\textcolor{black}{\begin{equation}
\label{energy}
\begin{array}{rl}
\displaystyle E-E_{bulk} & \displaystyle ={\cal M}\left(\sum 
^{N_{H}}_{j=1}\cosh h_{j}-2\sum ^{N_{S}}_{j=1}\cosh y_{j}-\sum 
^{M_{C}}_{j=1}\cosh c_{j}+
\sum _{j=1}^{M_{W}}(\cosh w_{j})_{II}-
\right.\\
 & \displaystyle \left. -\int ^{\infty 
}_{-\infty }\frac{dx}{2\pi }2Im\left[ \sinh (x+i\eta )\log (1+(-1)^{\delta 
}e^{iZ(x+i\eta )})\right] \right)
\end{array}
\end{equation}
\begin{equation}
\label{momentum}
\begin{array}{rl}
\displaystyle P & \displaystyle ={\cal M}\left(\sum ^{N_{H}}_{j=1}\sinh h_{j}-2
\sum^{N_{S}}_{j=1}\sinh y_{j}-\sum ^{M_{C}}_{j=1}\sinh c_{j}+
\sum _{j=1}^{M_{W}}(\sinh w_{j})_{II}-\right.\\
 & \displaystyle \left. -\int ^{\infty 
}_{-\infty }\frac{dx}{2\pi }2Im\left[ \cosh (x+i\eta )\log (1+(-1)^{\delta 
}e^{iZ(x+i\eta )})\right]\right) 
\end{array}
\end{equation}
Therefore, once the \( Z(\vartheta ) \) and the \( \vartheta _{j} \) positions
are calculated by the numerical iterative procedure illustrated above, it is
sufficient to plug their values into the expressions (\ref{energy}) and 
(\ref{momentum})
to get the values of these quantities for the given state with the wanted 
accuracy.
These values are} \textcolor{black}{\emph{numerical but exact,}} 
\textcolor{black}{in
the sense that, although there is lack of a closed formula for them, their 
computation
is done without any approximation other than the technical ones introduced by
the computer truncations.}

\subsection{Physical interpretation}

The limit procedure described in the previous section is mathematically 
consistent,
but the question is if from the physical point of view it describes a 
consistent
quantum theory and allows for a meaningful physical interpretation.

First of all, an important remark must be made about the allowed values for
the XXZ spin \( S \). It is clear from (\ref{spin-chain}) that on the lattice
only integer and nonnegative values can be taken into account for \( S \).
But on the continuum the definition of \( S \) is no more related to the Bethe
state (that is undefined), instead it is given implicitly by 
(\ref{counting-eq}).
Then, \emph{{}``a priori{}''}, there are no arguments that constrain its values
to be integers. As shown in \cite{noi PL2}, the half-integer choice for \( S \)
is necessary (and gives completely consistent results) to describe odd numbers
of particles.

At this point the following physical interpretation can be proposed. It will
be refined to describe the correspondence with particles. Also it will be 
supported
by many arguments that will be clarified in the following:

\begin{itemize}
\item the physical vacuum (Hamiltonian ground state) corresponds to absence of 
sources
(i.e. holes, complex roots,...); all the sources are excitations on this 
vacuum.
Indeed, it can be shown that they give positive contribution to the energy,
behaving as particle excitations on a vacuum state.
\item for \( \omega =0 \) and at the various values of \( S \) this theory 
describes
the sine-Gordon/massive Thirring model on a finite space of size \( L \); \( 
2S \)
is the topological charge and can take nonnegative integer values. 
\item for the values \begin{equation}
\label{omega}
\displaystyle \omega =\frac{k\pi }{s}\, ,\qquad k=1,...,q'-1
\end{equation}
 it describes the quantum restrictions of sine-Gordon model, i.e. the massive
integrable theory obtained perturbing the minimal models \( Vir(r,s) \) by
the operator \( \Phi _{(1,3)} \).
\item the real roots have disappeared from the counting in the continuum 
limit. They
actually become infinite in number and are taken into account by the integral
term, both in the NLIE and in the energy-momentum expression. They can be 
interpreted
as a sort of Dirac sea on which holes and complex roots build particle 
excitations.
Note that the presence of holes or complex roots, through the source term \( 
g(\vartheta |\vartheta _{j}) \)
distorts the Dirac sea too. This can be interpreted as vacuum polarization due
to the presence of particles.
\end{itemize}
Observe that it has been assumed that only nonnegative values of \( S \) are
required to describe the whole Hilbert space of the theory. Indeed the theory
is assumed charge-conjugation invariant. Then, negative topological charge 
states
have the same energy and momentum as their charge conjugate states. The 
assumption
that all the states can be described by the NLIE is absolutely not trivial.
A mathematical proof of this statement is not yet available, but a number of
specific cases supports this conjecture quite strongly.

\subsection{\textcolor{black}{The infrared limit of the NLIE and particle 
scattering}}

\textcolor{black}{The first task to understand the physics underlying the NLIE
we have found, is to identify the theory as a factorized scattering, 
reconstructing
the S-matrix.}

\textcolor{black}{In the IR limit \( l\rightarrow \infty  \), the convolution
term in (\ref{nlie-cont}) vanishes exponentially fast, so it can be dropped.
Consider first a state with \( N_{H} \) holes only and XXZ spin \( S=N_{H}/2 
\).}
\bigskip{}

\noindent \textcolor{black}{\begin{equation}
\label{IR-holes}
Z(\vartheta )=l\sinh \vartheta +\sum _{j=1}^{N_{H}}\chi (\vartheta 
-h_{j})\quad ,\quad Z(h_{j})=2\pi I_{j}
\end{equation}
Observe that the function \( \chi  \) can be written as \[
\chi (\vartheta )=-i\log S_{++}^{++}(\vartheta )\]
where \( S_{++}^{++}(\vartheta ) \) is the soliton-soliton scattering amplitude
in sine-Gordon theory \cite{zam79}, if the parameter \( p=\frac{\pi }{\gamma 
}-1 \)
is identified with the one introduced in sG theory \( p=\frac{\beta ^{2}}{8\pi 
-\beta ^{2}} \).
This fact is not surprising: eq.(\ref{IR-holes}) can be interpreted physically
as a quantization of momenta in a box, for a system of particles interacting
with \( \chi  \) as logarithm of the S-matrix. The fact that such quantization
is realized supports the interpretation of holes as solitons with rapidities
\( h_{j} \). This is further evidenced by considering the formula for the 
energy
in such a case. There too the integral term drops at IR, and the formula looks
like\[
E={\cal M}\sum _{j=1}^{N_{H}}\cosh h_{j}\]
which is the energy of \( N_{H} \) free particles of mass \( {\cal M} \).
The identification with the particular element \( S_{++}^{++} \) of the 
S-matrix
forces to give to these solitons a topological charge +1 each, which is 
consistent
with the interpretation that \( Q=2S \). To be correct, as \( 
S_{--}^{--}=S_{++}^{++} \)
an analogous interpretation is possible in terms of pure antisolitons, all with
charge --1, reflecting the charge conjugation invariance of the theory. }

Next state to consider is \textcolor{black}{two holes and a complex pair. After
some manipulation, the source terms can be arranged, thanks to some identities
satisfied by \( \chi  \) functions, in the form\[
Z(\vartheta _{i})=l\sinh \left( \vartheta _{i}\right) -i\log S_{-}(\vartheta 
_{i}-\vartheta _{j})=2\pi I_{j}\, \, \, ,\, \, \, i\, ,\, j=1,\, 2\]
where \[
S_{-}(\vartheta )=-\frac{\sinh \left( \frac{\vartheta +i\pi }{2p}\right) 
}{\sinh \left( \frac{\vartheta -i\pi }{2p}\right) }S^{++}_{++}(\vartheta )\]
which is the scattering amplitude of a soliton on an antisoliton in the 
parity-odd
channel. The quantum numbers \( I_{+},I_{-} \) of the two complex roots are
constrained to be \( I_{\pm }=\mp \frac{1}{2} \) for consistency of the IR
limit. This state has \( S=0 \). It must have topological charge 0, again 
consistent
with \( Q=2S \).}

It is known that there are two independent scattering amplitudes for soliton
- antisoliton scattering (usually they are presented as transmission and 
reflection,
here the alternative basis of parity even and odd channels is preferred). The
even channel is realized by the state with \textcolor{black}{two holes and a
selfconjugate root\[
Z(\vartheta _{i})=l\sinh \left( \vartheta _{i}\right) -i\log S_{+}(\vartheta 
_{i}-\vartheta _{j})=2\pi I_{j}\, \, \, ,\, \, \, i\, ,\, j=1,\, 2\]
where \[
S_{+}(\vartheta )=\frac{\cosh \left( \frac{\vartheta +i\pi }{2p}\right) 
}{\cosh \left( \frac{\vartheta -i\pi }{2p}\right) }S^{++}_{++}(\vartheta )\]
 which is the soliton-antisoliton amplitude in the parity-even channel.}

Thus, the whole soliton scattering S-matrix of sG theory has been 
reconstructed.
Note that, according to the counting equation (\ref{counting-eq}), in the 
repulsive
regime these 3 possibilities exhaust the possible configurations with \( 
N_{H}=2 \).

In the attractive regime one has also to consider the breather particles that
appear as soliton - antisoliton bound states. It turns out that the breathers
are represented by self-conjugate roots (1st breather) or by arrays of wide
roots (higher breathers). Consistency of quantization at the IR limit 
constrains
these arrays to appear only when \( p<1 \) decreases under the threshold point
\( p=1/k \) for the k-th breather. Similar considerations to the above lead
to the check that also the breather - soliton and breather - breather S-matrix
are correctly reproduced \cite{tani, noi NP2}.

\( N_{H}=1,3 \) states have been examined too \cite{noi PL2}. In particular
the one particle states give the correct description of a single soliton (if
\( \delta =1 \)) or massive Thirring fermion (if \( \delta =0 \)). In the
former case, the particle can be at rest has quantum number \( I_{h}=0 \).
In the latter, instead, the lowest possible quantum numbers, corresponding to
the lowest possible rapidities, are \( I_{h}=\pm \frac{1}{2} \). There is 
actually
a pair of states, in accordance with the fermion having two components.

\subsection{UV limit and vertex operators}

This identification gets more support if we consider, for each of the states
described above, the opposite \( l\rightarrow 0 \) limit, where we expect to
make contact with the UV limit of sG/mTh theory, i.e. with the \( c=1 \) CFT
described in section \ref{section:free_boson}. The UV calculations are usually
more difficult to perform than the IR ones, as they involve considering the
splitting of the NLIE at UV into two independent left and right parts (as it
should be in a good CFT!), the so called \emph{kink equations} and then 
expressing
the energy and momentum in closed form in terms of these, thanks to a lemma
presented in \cite{ddv 95}. For the details of this sort of manipulations the
reader is invited to consult the thesis \cite{tesi-Giovanni} where all the
calculations are done in detail. Here we present the main results and the 
physical
insight they imply.

A first important result is that the \( c=1 \) CFT quantum number \( m \)
(winding number), which is identified with the UV limit of the topological 
charge,
can be related unambiguously to the XXZ spin by \( \pm m=2S \). Of course,
the \( \pm  \) reflects the charge conjugation invariance of the theory. Then,
by examining the states we have already visited at IR, we can establish a 
bridge
between particle states and vertex operators of \( c=1 \) theory.

\begin{enumerate}
\item \textcolor{black}{The} \textbf{\textcolor{black}{vacuum}} 
\textcolor{black}{state
is the one with no holes of complex roots: only the sea of real roots is 
present.
There are two possible choices: \( \delta =0 \) or \( 1 \) (corresponding,
if reinterpreted as \( \alpha =0,\pi  \), to periodic and antiperiodic boundary
conditions respectively). The result of the UV calculation gives\[
\begin{array}{lll}
\rm {for}\, \delta =0:\quad  & \Delta ^{\pm }=0 & \, \, \mathrm{i}.e.\, \, 
\mathbb I\\
\rm {for}\, \delta =1:\quad  & \Delta ^{\pm }=\frac{1}{8R^{2}} & \, \, 
\mathrm{i}.e.\, \, V_{(\pm 1/2,0)}
\end{array}\]
i.e. the physical vacuum is the one with \( \delta =0 \). The other state 
belongs,
with reference to fig. \ref{4sectors.eps}, to the unphysical sector IV that
never contributes to make a local QFT at UV.}
\item \textcolor{black}{The} \textbf{\textcolor{black}{two-soliton}} 
\textcolor{black}{(two-hole)
state, with the minimal rapidity choice obtained with the lowest possible (in
absolute value) quantum numbers, gives}

\begin{enumerate}
\item \textcolor{black}{for \( \delta =0 \) and \( I_{1}=-I_{2}=\frac{1}{2} \) 
\( \Longrightarrow  \)
\( \Delta ^{\pm }=\frac{R^{2}}{2}\, \, \mathrm{i}.e.\, \, V_{(0,2)} \). }
\item \textcolor{black}{for \( \delta =1 \) and \( I_{1}=-I_{2}=1 \) \( 
\Longrightarrow  \)
a \( V_{(\pm 1/2,2)} \) descendent, not in UV sG spectrum, as it also belongs
to sector IV.}
\end{enumerate}
\item \textcolor{black}{The} \textbf{\textcolor{black}{symmetric 
soliton-antisoliton}}
\textcolor{black}{state (two holes and a self-conjugate root), always with the
minimal rapidity choice} \textcolor{black}{}\\
\textcolor{black}{~~\( \delta =1 \), \( I_{1}=-I_{2}=1 \) and \( I_{c}^{\pm 
}=0 \)
\( \Longrightarrow  \) \( \Delta ^{\pm }=\frac{1}{2R^{2}}\, \, \mathrm{i}.e.\, 
\, V_{(\pm 1,0)} \)}
\item \textcolor{black}{The} \textbf{\textcolor{black}{antisymmetric 
soliton-antisoliton}}
\textcolor{black}{state (two holes and a complex pair) }\\
\textcolor{black}{~~\( \delta =0 \), \( 
I_{1}=I_{c}^{-}=-I_{2}=-I_{c}^{+}=\frac{1}{2} \)
\( \Longrightarrow  \) \( \Delta ^{\pm }=\frac{1}{2R^{2}}\, \, \mathrm{i}.e.\, 
\, V_{(\pm 1,0)} \)}\\
\textcolor{black}{It is obvious that these last two give two linearly 
independent
combinations of the operators \( V_{(\pm 1,0)} \), one with even, the other
with odd parity.}
\item \textcolor{black}{The} \textbf{\textcolor{black}{one hole}} 
\textcolor{black}{state}
\textbf{\textcolor{black}{}}\textcolor{black}{with} 
\textbf{\textcolor{black}{}}\textcolor{black}{\( I=0 \),
\( \delta =1 \) \( \Longrightarrow  \) \( \Delta ^{\pm }=\frac{1}{8R^{2}} \)
i.e. the vertex operator \( V_{(0,1)} \), belonging to sector II, thus giving
more support to the 1 soliton interpretation. For \( \delta =0 \) there are
two minimal rapidity states with \( I=\pm \frac{1}{2} \). They are identified
with the operators \( V_{(\pm 1/2,\pm 1)} \). As these states belong to sector
III, they are of fermionic nature and actually one identifies them with the
components of the Thirring fermion.}
\item \textcolor{black}{The} \textbf{\textcolor{black}{three holes}} 
\textcolor{black}{state
with \( I_{1}=1 \), \( I_{2}=0 \), \( I_{3}=-1 \) \( \Longrightarrow  \)
\( V_{(0,3)} \). Higher values of quantum numbers give \( V_{(n,3)} \), \( 
n\in \mathbb {Z} \) }
\item \textcolor{black}{The} \textbf{\textcolor{black}{three holes}} 
\textcolor{black}{}\textbf{\textcolor{black}{and
close pair}} \textcolor{black}{state gives the first nonchiral descendent of
\( V_{(0,1)} \).}
\end{enumerate}
\textcolor{black}{These examples, taken all together, suggest a quantization
rule (i.e a consistent choice of \( \delta ) \) to obtain states which are
present in the UV spectrum of sine-Gordon or massive Thirring theory. It looks
like\begin{equation}
\label{regola_d'oro}
\begin{array}{cc}
Q+\delta +M_{sc}\in 2\mathbb Z & \quad \rm {for\, Sine-Gordon}\\
\delta +M_{sc}\in 2\mathbb Z & \quad \rm {for\, massive\, Thirring}
\end{array}
\end{equation}
where \( M_{sc} \) is the number of selfconjugate roots. This selects sector
I and II for sine-Gordon states, and sectors I and III for Thirring ones, as
it should be from the discussion of sections \ref{section:free_boson} and 
\ref{section:sine-gordon}.
It always excludes sector IV, that does not contain local operators.} All this
is in accordance with the correct interpretation of Coleman equivalence of 
Sine-Gordon
and Thirring models: even topological charge sectors are identical and the 
difference
of the two models shows up only in the odd topological charge sectors, for 
which
the content of Thirring must be fermionic while that of Sine-Gordon must be
bosonic.

To conclude these remarks, we briefly comment about special objects that were
introduced in the classification of roots but never used later. Recall their
definition: they are roots or holes \( y_{i} \) having \( Z'(y_{i})<0 \).
Now, the function \( Z \) is globally monotonically increasing. Indeed its
asymptotics for \( \vartheta \rightarrow \pm \infty  \) are dominated by the
term \( l\sinh \vartheta  \) which is obviously monotonically increasing. Also,
for \( l \) large, this term dominates. Therefore at IR the function \( Z \)
is surely monotonic and no special objects can appear. This is why we have not
considered them in the classification of states above. However, it can happen
that at some critical value \( l_{crit} \) of \( l \), coming back from IR
towards UV, the convergence of the iterative procedure breaks down, thus 
revealing
that some singularity has been encountered. For the scaling function to be 
consistently
analytically continued after this singularity to reach the UV regime, one needs
to introduce modifications to the NLIE creating exactly the contributions that
we have called special objects. A more careful analysis reveals that these 
singularities
are produced by the logarithm in the convolution term going off its fundamental
branch. A detailed treatment of these objects can be found in 
\cite{tesi-Giovanni}.

\subsection{Twisted NLIE and minimal models}

It is a well known fact \cite{reshetikhin_smirnov} that the perturbation of
the Virasoro minimal model \( Vir(r,s) \) by its relevant primary operator
\( \Phi _{(1,3)} \) is integrable and is described by an RSOS restriction of
sine-Gordon theory with \begin{equation}
\label{p-rs}
\displaystyle p=\frac{r}{s-r}\, \, .
\end{equation}
 We will use for this model the shorthand notation \( Vir(r,s)+\Phi _{(1,3)} 
\).
Al. Zamolodchikov has put forward the idea of modifying sine-Gordon NLIE by
a twist \( \alpha  \) \cite{polymer} to deal with conformal minimal models.
It is to allow this sort of approach that since the beginning we have 
considered
the light-cone lattice construction with a twist \( \omega  \) that induced
a twist \( \alpha  \) in the NLIE. All the physics of sG/mTh model can be done
with \( \alpha =0 \). Now we explore the situation where \( \alpha  \) can
take well chosen nonzero values implementing the reduction and allowing 
therefore
the NLIE to describe scaling functions of perturbed minimal models too.

Looking first at the ground state, that in analogy with sine-Gordon is expected
to be a sea of real roots, the source in NLIE is put to zero and we choose 
half-integer
quantization with \( \delta =0 \). The ultraviolet limit of the scaling 
functions
can be computed and gives \begin{equation}
\label{alpha_{v}acuum}
\displaystyle \tilde{c}=1-\frac{6p}{p+1}\left( \frac{\alpha }{\pi }\right) ^{2}
\end{equation}
 Only in the unitary models \( \tilde{c} \) is the Virasoro central charge.
Using the rule suggested in (\ref{omega}) for \( \omega  \) and the expression
(\ref{alfa}) for \( \alpha  \) gives: \( \alpha =\pi /r \). Then \[
\tilde{c}=1-\frac{6}{rs}\, \, ,\]
 which is exactly the \emph{effective central charge} \( \tilde{c}=c-24\Delta 
_{min} \)
of the minimal model \( Vir(r,s) \). Therefore one can expect that the twisted
equation describes the ground state of the model \( Vir(r,s)+\Phi _{(1,3)} \).
In fact, Fioravanti et al. \cite{fioravanti} calculated these scaling functions
for the unitary case \( s=r+1 \) and showed that they match perfectly with
the TBA predictions already available. Moreover, choosing the following values
for the twist \begin{equation}
\label{twists}
\displaystyle \alpha =\pm \frac{k\pi }{r}\, \, ,\, \, k=1\ldots r-1\, \, 
\end{equation}
 they obtained the conformal weights of the operators \( \Phi _{(k,k)}\, \, 
,\, \, k=1\ldots r-1 \)
in the UV limit (the sign choice is just a matter of convention). In our 
notation,
\( \Phi _{(q,q')} \) denotes the primary field with conformal weights 
\begin{equation}
\label{Kac_{f}ormula}
\displaystyle \Delta =\bar{\Delta }=\frac{(qs-q'r)^{2}-(s-r)^{2}}{4sr}\, \, .
\end{equation}
 The models \( Vir(r,s)+\Phi _{(1,3)} \) have exactly \( r-1 \) ground states.
In fact, one can see from the fusion rules that the matrix of the operator \( 
\Phi _{(1,3)} \)
is block diagonal with exactly \( r-1 \) blocks in the Hilbert space made up
of states with the same left and right primary weights. In each of these 
blocks,
there is exactly one ground state and for the unitary series \( s=r+1 \), it
was conjectured in \cite{kl-me2} that their UV limits are the states 
corresponding
to \( \Phi _{(k,k)} \). One can check that in the general non-unitary case
the twists (\ref{twists}) correspond in the UV limit to the lowest dimension
operators among each of the \( r-1 \) different blocks of primaries (see 
explicit
examples later).

These ground states are degenerate in infinite volume, but for finite \( l \)
they split; their gaps decay exponentially as \( l\, \rightarrow \, \infty  \).
In the unitary case, they were first analyzed in the context of the NLIE in
\cite{fioravanti} where it was shown that the NLIE predictions perfectly match
with the TBA results already available for the unitary series.

Ground states for non-unitary models have been treated in \cite{noi NP2}. The
simplest example is the scaling Lee-Yang model \( Vir(2,5)+\Phi _{(1,3)} \),
for which \[
M_{B}=2{\mathcal{M}}\sin \frac{\pi p}{2}=\sqrt{3}{\mathcal{M}}\]
 is the mass of the fundamental particle of the Lee-Yang model (this is more
natural here than using the mass \( {\mathcal{M}} \) of the soliton of the
unrestricted sine-Gordon model as a scale, since the soliton disappears 
entirely
from the spectrum after RSOS restriction). Call \[
l_{B}=M_{B}L=\sqrt{3}l\, \]
 where \( l \) is the variable appearing in NLIE. There is only one independent
value of the twist \[
\alpha =\frac{\pi }{2}\, \, .\]

There is only one ground state in this model, which corresponds to the primary
field with conformal weights \[
\Delta =\bar{\Delta }=-\frac{1}{5}\, \, ,\]
 All the models of the class \( Vir(2,2n+1)+\Phi _{(1,3)} \) have only one
ground state. For models with two ground states, we can take a look at \( 
Vir(3,5) \)
(\( Vir(3,7) \) was also taken into account in \cite{noi NP2}). For \( 
Vir(3,5) \)
the ultraviolet spectrum is defined by the following Kac table, where the 
weight
of the field \( \Phi _{(k,l)} \) is found in the \( k \)-th row and \( l \)-th
column.
\vspace{1mm}

{\centering \begin{tabular}{|c|c|c|c|}
\hline 
0   & --1/20 & 1/5 & 3/4\\
\hline 
3/4 & 1/5 & --1/20 & 0\\
\hline 
\end{tabular}\par}

The two blocks of the perturbing operator \( \Phi _{(1,3)} \) are defined by
the fields \( \{\Phi _{(1,2)}\, ,\, \Phi _{(1,4)}\} \) and \( \{\Phi 
_{(1,1)}\, ,\, \Phi _{(1,3)}\} \),
respectively. The ground states correspond in the UV to the operators \( \Phi 
_{(1,2)} \)
and \( \Phi _{(1,1)} \), as can be checked directly using formulae 
(\ref{alpha_{v}acuum}),
(\ref{p-rs}) and (\ref{twists}).

Now we turn our attention to excited states over these minimal model vacua.
We restrict ourselves to the case of neutral (i.e. \( S=0 \)) states. Even
for states with a zero charge the relation between \( \alpha  \) and \( \omega 
 \)
is highly nontrivial.

It is quite easy to show that choosing the value of \( \omega  \) as \[
\omega =\frac{k\pi }{p+1}\]
 where \( k \) is integer, we can reproduce all the required values of \( 
\alpha  \)
listed in equation (\ref{twists}). The twisted lattice Bethe Ansatz was 
analyzed
by de Vega and Giacomini in \cite{omega-twist}. On the lattice, passing from
the sine-Gordon model to the perturbed Virasoro model amounts to going from
the six-vertex model to a lattice RSOS model. In \cite{omega-twist} it was
shown that to obtain all the states of the RSOS model it is necessary to take
all the twists \[
\omega =\frac{k\pi }{p+1}\, \, \bmod \, \, \pi \]
 into account. The fact that not all these twists correspond to inequivalent
values of \( \alpha  \) and so to different physical states is a consequence
of the RSOS truncation.

We remark that the parameter \( \alpha  \) drops out of the second 
determination
of \( Z \) in the attractive regime. This is important because as a consequence
the IR asymptotics of the breather states do not depend on \( \alpha  \) and
so the \( S \)-matrices involving breathers are unchanged. In fact, scattering
amplitudes between solitons and breathers remain unchanged too. This matches
with the fact that the RSOS restriction from sine-Gordon theory to perturbed
minimal models does not modify scattering amplitudes that involve two breathers
or a breather and soliton \cite{reshetikhin_smirnov}.

Let us start by examining the scaling Lee-Yang model \( Vir(2,5)+\Phi _{(1,3)} 
\).
Since we have a single ground state, there are no kinks in the spectrum. We
fix the value of \( \alpha  \) as above, so we still have a freedom of choosing
\( \delta  \). This can be done by matching to the UV dimensions: if for a
certain state we choose the wrong value of \( \delta  \), we find a conformal
dimension that is not present in the Kac table of the model.

The excited states are multi-particle states of the first breather of the 
corresponding
unrestricted sine-Gordon model, which has \( p=\frac{2}{3} \). One can 
calculate
the state containing one particle at rest. It turns out that as we decrease
\( l \), the self-conjugate root starts moving to the right. It does not remain
in the middle like in the \( \alpha =0 \) case, which is to be expected since
for nonzero \( \alpha  \) we have no left/right symmetry. However, the total
momentum of the state still remains zero due to a contribution from the 
integral
term in momentum equation (\ref{momentum}). We have here an example of the
phenomenon of the appearance of the special root and its two accompanying 
holes:
the numerical iteration breaks down at around \( l_{crit}=2.5 \). For \( 
l<l_{crit} \)
the NLIE must be corrected by the introduction of this special root and the
two accompanying holes (which are not special). They must be taken into account
to compute the UV limit correctly. Notice that the two holes thus introduced
do not have any counterpart at IR and it is not possible to give them any 
interpretation
as particles. Their quantum numbers cannot be freely chosen. They are 
constrained
to take the same value of the quantum number of the special root, which in turn
is fixed by the constraints of the problem.

We know that the Lee-Yang model contains only two primary fields, the identity
\( \mathbb I \) and the field \( \varphi  \) with left/right conformal weights
--1/5. In fact, the ground state of the massive model corresponds to \( 
\varphi  \)
in the UV limit. For the one particle state, it turns out that the special root
and one of the holes move to the left together with the self-conjugate root,
while the other hole moves to the right. This gives, after the {}``kink{}''
NLIE calculations are done, \( \Delta =\bar{\Delta }=0 \), i.e. the identity
operator \( \mathbb I \), as expected from TCSA in \cite{yurov-zam}.

Let us look now at moving breathers. If the self-conjugate root has Bethe 
quantum
number \( I=1 \), the corresponding state will have momentum quantum number
\( 1 \), i.e. \[
P=\frac{2\pi }{R}\, \, ,\]
 and in the UV \( \Delta -\bar{\Delta }=1 \). No special root appears here.
The reason is that the self-conjugate root moves to the left and the real part
of its position \( \vartheta  \) is given to leading order by \[
\sinh (\Re e\, \vartheta )\, \sim \, -\frac{2\pi I}{l_{B}}\, \, .\]
 As a result, the contribution to the derivative of \( Z \) from the \( l\sinh 
\vartheta  \)
term remains finite when \( l\, \rightarrow \, 0 \). In the previous example
of the particle at rest the left-moving nature of the self-conjugate root when
\( I=0 \) does not prevent the occurrence of the breakdown in the iteration
scheme: since its Bethe quantum number is zero, it does not move fast enough
to the left in order to balance the negative contribution to derivative of \( 
Z \)
coming from the self-conjugate root source. At the moment we have no way of
predicting analytically whether or not there will be specials in the UV limit:
we just use the numerical results to establish the configuration for the 
evaluation
of UV weights, supplemented with a study of the self-consistency of the 
solution
of the kink equation. The UV dimensions for the moving breather turn out to
correspond to the state \( L_{-1}\varphi  \).

One can similarly compute the UV dimensions for some other excited states. For
example, the two-particle states with half-integer Bethe quantum numbers \( 
I_{1}>0,\, \, I_{2}<0 \)
for the two self-conjugate roots are found to have \[
\Delta =-\frac{1}{5}+I_{1}+\frac{1}{2}\, \, ,\, \, \bar{\Delta 
}=-\frac{1}{5}-I_{2}+\frac{1}{2}\, \, ,\]
 which shows that they correspond in the UV to descendent states of \( \varphi 
 \).
The first such state with quantum numbers \[
I_{1}=\frac{1}{2}\, \, ,\, \, I_{2}=-\frac{1}{2}\]
 corresponds to \( L_{-1}\bar{L}_{-1}\varphi  \).

The lowest lying three-particle state of zero momentum, with Bethe quantum 
numbers
\( (-1,0,1) \) corresponds to the left/right symmetric second descendent of
the identity field, i.e. to the field \( T\bar{T} \), where \( T \) denotes
the energy-momentum tensor. This is very interesting, since from experience
with NLIE UV calculations one would naively expect this to be a first 
descendent
(descendent numbers are usually linked to the sum of Bethe quantum numbers of
left/right moving particles and this state is the lowest possible descendent
of the identity \( \mathbb I \)). However, the field \( 
L_{-1}\bar{L}_{-1}\mathbb I \)
is well-known to be a null field in any conformal field theory. It seems from
this example that NLIE is clever enough to avoid null vectors in minimal 
models,
but of course more investigation is needed in this respect.

The above correspondences are confirmed by comparing to TCSA data (see the 
wonderful
figures in \cite{yurov-zam}). In general, one can establish the rule that 
states
with odd number of particles must be quantized by integers (\( \delta =1 \)),
while those containing even number of particles must be quantized by 
half-integers
(\( \delta =0 \)) in order to reproduce correctly the spectrum of the scaling
Lee-Yang model.

We conducted similar studies for the models \( Vir(2,7)+\Phi _{(1,3)} \) and
\( Vir(2,9)+\Phi _{(1,3)} \) and found similarly good agreement with TCSA data.
For the first one-particle state of the model \( Vir(2,7)+\Phi _{(1,3)} \)
we also checked our results against the TBA data in the numerical tables of
\cite{dorey-tateo} and found agreement with the TBA results.

Given the choice of \( \alpha  \) above, the correct rule of quantization in
all of the models \( Vir(2,2n+1)+\Phi _{(1,3)} \) is \[
\delta +M_{sc}\in 2\mathbb Z\]
 where \( M_{sc} \) is the number of self-conjugate roots in the source 
corresponding
to the state. This is exactly the same rule as the one established for pure
sine-Gordon theory in \cite{noi NP}. In the presence of the twist, such a rule
of course has meaning only together with a definite convention for the choice
of \( \alpha  \).

It is interesting to note that in the case of \( Vir(3,n) \) models, all the
neutral states must come in two copies, since they can be built on top of 
either
of the two ground states. We take the example of the \( Vir(3,7) \) model and
the states corresponding to a breather at rest. We have \( p=\frac{3}{4} \),
but now there are two inequivalent values for the twist \[
\alpha =\frac{\pi }{3}\, \, ,\, \, \frac{2\pi }{3}\, .\]
 When \( \alpha =0 \), we can estimate the critical value of \( l \) where
specials appear to be \( l_{crit}=5.23 \). In this case, the twist helps a
bit, because it makes the self-conjugate root a left mover; it is intuitively
clear that the bigger the twist, the more it lowers the eventual value of \( 
l_{crit} \).
From TCSA data one can identify that breather \#1 is really the one-particle
state in the sector of ground state \#1 (\( \alpha =\pi /3 \)), while breather
\#2 is in the sector over ground state \#2 (\( \alpha =2\pi /3 \)).

A direct calculation of the conformal weights gives \( \frac{3}{28} \) for
breather \#1 and \( 0 \) for breather \#2, which are in complete agreement
with TCSA data. We also checked the two different states containing two 
breathers
with Bethe quantum numbers \( I_{1}=\frac{1}{2}\, \, ,\, \, 
I_{2}=-\frac{1}{2}\, \, , \)
and found an equally excellent numerical agreement with TCSA. Just like in the
case of sine-Gordon and scaling Lee-Yang model, for these states one can 
continue
the iteration of the NLIE down to any small value of \( l \), although at the
expense of a growing number of necessary iterations to achieve the prescribed
precision.

\section{Conclusions\label{section:conclusione}}

We have shown how the nonlinear integral equation deduced from the light cone
lattice model of \cite{ddv 87} describes exactly the Finite Size Effects of
the sine-Gordon/massive Thirring theory, taking into account excited states
too. The most important results are summarized as follows:

\begin{enumerate}
\item By examining the infrared limits of the equation it has been shown that 
it leads
to the correct two-particle S-matrices for both scattering states and bound
states;
\item By computing the UV conformal weights from the NLIE we have shown that 
the energy-momentum
spectrum is consistent with the UV one of sG/mTh theory only if we choose the
parameter \( \delta  \) (i.e. the quantization rule) as indicated in 
(\ref{regola_d'oro}). 
\item The predictions of the NLIE have been verified by comparing them to 
results
coming from the TCSA approach (for sG/mTh). 
\item The framework required to deal with minimal models perturbed by \( \Phi 
_{(1,3)} \)
is built up. Many examples and numerical/analytical checks are given for the
ground state in the unitary \cite{fioravanti} and non unitary cases. 
\item All the conformal dimensions can be reproduced (Kac table) with a 
convenient
choice of the twist (\ref{twists}). The particular relation suggested in 
\cite{zinn-justin}
is not enough to describe the whole spectrum. 
\item The IR computations reproduce correctly the S-matrix of perturbed 
minimal models
(at list in the cases not involving kinks), because the \( p<1 \) second 
determination
drops the twist. 
\item Numerical calculations (TCSA) of concrete examples and comparison with 
TBA where
possible, give a strong evidence for the correctness of the energy levels 
derived
from the twisted NLIE for excited states, both for the full sG/mTh model and
for the Restricted models.
\end{enumerate}
The understanding of {}``(twisted) sine-Gordon NLIE{}'' and of the finite
size behaviour of the continuum theory defined from the NLIE (\ref{nlie-cont})
is not quite complete. Indeed some open questions have not an answer yet and
further interesting developments may be forecast. We close this review by 
pointing
to possible interesting questions that need further investigation.

\begin{enumerate}
\item Is the set of scaling functions provided by the NLIE complete i.e. can 
we find
to every sG/mTh or minimal model state a solution of the NLIE describing its
finite volume behaviour? The main difficulty in this {}``counting problem{}''
is that the structure of the solutions is highly dependent on the value of the
coupling constant -- consider e.g. the appearance of special sources. 
\item The multi-kink states characteristic of the perturbed minimal models 
(except
for the series \( Vir(2,2n+1) \)) have been omitted in the analysis performed
in the last section. Although the general treatment is valid for such states
too, a detailed description is far more complicated than for states which 
contain
only breathers and is left open to further studies. 
\item There is also an unresolved technical difficulty, namely that the source 
configuration
of the NLIE may change as we vary the volume parameter \( l \). Typically what
happens is that while the counting function \( Z \) is monotonic on the real
axis for large volume, this may change as we lower the value of \( l \) and
so-called special sources (and accompanying holes) may appear. We do not as
yet have any consistent and tractable numerical iteration scheme to handle this
situation, although the analytic UV calculations and intuitive arguments show
that the appearance of these terms in the NLIE is consistent with all 
expectations
coming from the known properties of perturbed CFT. In addition, in the range
of \( l \) where we can iterate the NLIE without difficulty, our numerical
results show perfect agreement with TCSA. We want to emphasize that these 
transitions
are not physical: the counting function \( Z \) and the energy of the state
are expected to vary analytically with the volume. It is just their description
by the NLIE with requires a modification of source terms. As it was pointed
out also in \cite{ddv 97, noi NP}, the whole issue is related to the choice
of the branch of the logarithmic term in the NLIE (\ref{nlie-cont}). 
\item The problem pointed out at the previous point is very similar to the 
behaviour
of singularities encountered in the study of the analytic continuation of the
TBA equation \cite{dorey-tateo} and we can hope that establishing a closer
link between the two approaches can help to clarify the situation. From the
form of the source terms in the NLIE it seems likely that the excited state
equations can be obtained by an analytic continuation procedure analogous to
the one used in TBA \cite{dorey-tateo} to obtain the excited state TBA 
equations.
Certain features of the arrangement of the complex roots in the attractive 
regime
and their behaviour at breather thresholds also point into this direction. This
is an interesting question to investigate, because it can shed light on the
organization of the space of states and can lead closer to solving the counting
problem described above. 
\item Even if on the lattice the Bethe vectors given in (\ref{bethe}) 
completely
describe the Hilbert space of the XXZ chain, the form of eigenvectors for 
continuum
energy and momentum is completely unknown. This fact reflects the unresolved
question of the form of energy and momentum eigenvectors for the minkowskian
sine-Gordon theory (the Faddeev-Zamolodchikov algebra is only a 
phenomenological
picture). Also the determination of correlation functions is in an early stage,
although in recent publications \cite{vev-sG} explicit expressions for the
field \( \left\langle e^{ia\varphi }\right\rangle  \) and its descendent are
given. 
\item The extension of light-cone approach and NLIE to other QFT models is in 
progress.
An interesting extension is the description of finite volume spectrum of 
Zhiber-Mikhailov-Shabat
model (imaginary Bullough-Dodd). A first suggestion in this direction, even
if is obtained in a completely different approach, recently appeared in the
\cite{dorey-tateo_ultimo}, for the ground state. Early attempts to get the
excited states can be found in \cite{tesi-Marco}. Extensions to vertex models
based on higher rank simply-laced Lie Algebras have also been investigated 
\cite{mariottini,zinn-justin}
and should correspond -- a detailed analysis has not been done yet -- to the
solitonic (imaginary coupling) affine Toda Field Theories.
\end{enumerate}
The second and fourth points are important because their investigation may lead
closer to understanding the relation between the TBA and the NLIE approaches.
It is quite likely that establishing a connection between the two methods would
facilitate the development of both and may point to some common underlying 
structure.

As a final remark, let us mention an application of the NLIE to a chemical 
compound,
i.e. {}``copper Benzoate{}'' (\( 
\textrm{Cu}(\textrm{C}_{6}\textrm{D}_{5}\textrm{COO})_{2}\cdot 
3\textrm{D}_{2}\textrm{O} \)).
Its specific heat has been computed in \cite{copper_benzoate} using the vacuum
untwisted (\( \alpha =0 \)) NLIE, where the finite size \( L \) is the inverse
temperature \( L=(k_{B}T)^{-1} \). This corresponds to do equilibrium 
thermodynamics
of sine-Gordon model. It is curious to notice that in that context the NLIE
is called Thermal Bethe Ansatz, emphasizing the well known relation between
statistical mechanics and QFT. Exactly in a similar contest (Heisenberg 
ferromagnet
model) the original Bethe Ansatz \cite{bethe} was born.

\section*{Acknowledgements}

First of all I have to express my delight to have been invited to give these
lectures in one of the most beautiful towns of Europe, full of history and 
character,
by a group of enthusiast and experienced Field Theorists with whom it has been
enjoyable to have exchange of ideas and fruitful discussions. The kindness and
hospitality of Laszlo Palla, Zoltan Horvath and Zoltan Banjok and their 
efficiency
in organization is unforgettable.

Then I have to thank my collaborators without whom many of the results exposed
here would have never seen the light. Gabor Tákacs and Giovanni Feverati (whose
Ph.D thesis has been an essential help in writing parts of this review) are
particularly thanked for the splendid and stimulating two years spent together
in Bologna, during which much of the work on NLIE has been done. Ennio 
Quattrini,
Andrea Mariottini and Davide Fioravanti, that contributed to the first paper
of our group on excited states in NLIE, are also warmly acknowledged.

I have to thank Roberto Tateo and Patrick Dorey for the long friendship and
fruitful discussions through years that contributed to the general 
comprehension
of the subject and for the collaboration on various related subjects.

And a final and particular \emph{thank you} goes to my wife Irene and my 
children
Simone, Monica and Fabio that with their joy and love have supported me (in
all senses...) through this work.

\end{document}